\documentclass[fleqn,usenatbib]{mnras}

\usepackage{newtxtext,newtxmath}
\usepackage[T1]{fontenc}
\usepackage{ae,aecompl}

\usepackage{graphicx}	% Including figure files
\usepackage{amsmath}	% Advanced maths commands
\newcommand{\HI}{H\,{\sc i}}

\newcommand{\Msun}{~M$_{\odot}$}
\newcommand{\MMsun}{M$_{\odot}$}

\newcommand{\kms}{~km\,s$^{-1}$}

\newcommand{\zph}{$z_{\rm ph}$}

\usepackage{xcolor}

\title[Connecting cluster and galaxy merger shocks]{MeerKAT discovery of a double radio relic and odd radio circle: connecting cluster and galaxy merger shocks}

\author[Koribalski et al.]{B\"arbel S. Koribalski,$^{1,2}$\thanks{E-mail: Baerbel.Koribalski@csiro.au} Angie Veronica,$^{3}$ Klaus Dolag,$^{4,5}$ Thomas H. Reiprich,$^{3}$ \newauthor Marcus Br\"uggen,$^{6}$ Ian Heywood,$^{7,8,9}$ Heinz Andernach,$^{10,11}$  Ralf-J\"urgen Dettmar,$^{12}$ \newauthor Matthias Hoeft,$^{11}$ Xiaoyuan Zhang,$^{13}$ Esra Bulbul,$^{13}$ Christian Garrel,$^{13}$ Gyula I.G. \newauthor J\'ozsa,$^{14,8}$ and Jayanne English$^{15}$  \\
$^{1}$Australia Telescope National Facility, CSIRO Astronomy and Space Science, P.O. Box 76, Epping, NSW 1710, Australia \\
$^{2}$School of Science, Western Sydney University, Locked Bag 1797, Penrith, NSW 2751, Australia \\
$^{3}$Argelander-Institut f\"ur Astronomie (AIfA), Universit\"at Bonn, Auf dem H\"ugel 71, 53121 Bonn, Germany \\
$^{4}$Universit\"ats-Sternwarte, Fakult\"at f\"ur Physik, Ludwig-Maximilians-Universit\"at M\"unchen, Scheinerstr.1, 81679 M\"unchen, Germany \\
$^{4}$Max-Planck-Institut f\"ur Astrophysik, Karl-Schwarzschild-Str. 1, 85741 Garching, Germany \\
$^{6}$Hamburger Sternwarte, University of Hamburg, Gojenbergsweg 112, 21029 Hamburg, Germany \\
$^{7}$Astrophysics, Department of Physics, University of Oxford, Keble Road, Oxford, OX1 3RH, UK \\
$^{8}$Department of Physics and Electronics, Rhodes University, PO Box 94, Makhanda, 6140, South Africa \\
$^{9}$South African Radio Astronomy Observatory, 2 Fir Street, Observatory 7925, South Africa \\
$^{10}$Departmento de Astronom\'ia, DCNE, Universidad de Guanajuato, Callej\'on de Jalisco s/n,
Guanajuato, C.P. 36023, GTO, Mexico \\
$^{11}$Th\"uringer Landessternwarte, Sternwarte 5, 07778 Tautenburg, Germany \\
$^{12}$Ruhr University Bochum, Faculty of Physics and Astronomy, Astronomical Institute (AIRUB), 44780 Bochum, Germany \\
$^{13}$Max Planck Institute for Extraterrestrial Physics, Giessenbachstrasse 1, 85748 Garching, Germany \\
$^{14}$Max-Planck-Institut f\"ur Radioastronomie, Auf dem H\"ugel 69, 53121, Bonn, Germany \\
$^{15}$Department of Physics and Astronomy, University of Manitoba, Winnipeg, Manitoba R3T 2N2, Canada  
}

\date{Accepted XXX. Received YYY; in original form ZZZ}

\pubyear{2024}

\begin{document}
\label{firstpage}
\pagerange{\pageref{firstpage}--\pageref{lastpage}}
\maketitle

\begin{abstract}
We present the serendipitous discovery of (1) a large double radio relic associated with the galaxy cluster PSZ2~G277.93+12.34 and (2) a new odd radio circle, ORC~J1027--4422, both found in the same deep MeerKAT 1.3~GHz wide-band radio continuum image. The angular separation of the two arc-shaped cluster relics is $\sim$16\arcmin\ or $\sim$2.6~Mpc for a cluster redshift of $z \approx 0.158$. The thin southern relic, which shows several ridges/shocks including one possibly moving inwards, has a linear extent of $\sim$1.64~Mpc. In contrast, the northern relic is about twice as wide, twice as bright, but only has a largest linear size of $\sim$0.66~Mpc. Complementary SRG/eROSITA X-ray images reveal extended emission from hot intracluster gas between the two relics and around the narrow-angle tail (NAT) radio galaxy PMN J1033--4335 ($z \approx 0.153$) located just east of the northern relic. The radio morphologies of the NAT galaxy and the northern relic, which are also detected with the Australian Square Kilometer Array Pathfinder (ASKAP) at 888~MHz, suggest both are moving in the same outward direction. The discovery of ORC~J1027--4422 in a different part of the same MeerKAT image makes it the 4th known single ORC. It has a diameter of $\sim$90\arcsec\ corresponding to 400~kpc at a tentative redshift of $z \approx 0.3$ and remains undetected in X-ray emission. Supported by simulations, we discuss similarities between outward moving galaxy and cluster merger shocks as the formation mechanisms for ORCs and radio relics, respectively.

% The abstract should briefly describe the aims, methods, and main results of the paper. It should be a single paragraph not more than 250 words (200 words for Letters). No references should appear in the abstract.
\end{abstract}

% Select between one and six entries from the list of approved keywords. - Don't make up new ones.

%%%%%%%%%%%%%%%%%%%%%%%%%%%%%%%%%%%%%%%%%%%%%%%%%%

%%%%%%%%%%%%%%%%% BODY OF PAPER %%%%%%%%%%%%%%%%%%

\section{Introduction} 
\label{sec:intro}
Radio relics are diffuse, steep-spectrum radio synchrotron sources that typically occur in the form of single or double symmetric arcs at the peripheries of galaxy clusters \citep[e.g.,][]{Roettgering1997}. They can be explained by shock waves driven into the intracluster medium (ICM) by cluster mergers \cite[e.g.,][]{vanWeeren19, Thoelken2018, 2020MNRAS.493.2306B, Ghirardini2021, Brueggen2021, Hoang2022, Boess2023, Jones2023}. Their large size, arc-like radio morphology, high degree of polarization, and spectral index distribution make them very distinct from other extended radio sources. Galaxy clusters that host double relics are particularly interesting (and rare), because their geometry suggests that the merger is proceeding close to the plane of the sky. Odd radio circles \citep[ORCs,][]{Norris2021,Koribalski2021} recently discovered around massive early-type galaxies are similar in morphology to double relics around clusters but smaller with typical sizes of 300 -- 500~kpc. Using high-resolution cosmological simulations, \citet{Dolag2023} find that outwards moving merger-induced shock waves resembling ORCs occasionally occur during the formation of massive early-type galaxies. In both galaxy cluster and groups, merger-induced shock waves get re-energised by their propagation through the intracluster/group medium, and become observable as steep spectrum arc-shaped radio sources. \\

There is significant interest in radio relics because they are very efficient particle accelerators \citep{Botteon18}, sites of relatively strong magnetic fields \citep{Rajpurohit2022}, and indicators of the dynamical state of galaxy clusters \citep{Zhang2023}. Similarly, the formation of ORCs is not yet understood and a handful of models have been proposed. Finding more ORCs, which are extremely rare (only three single ORCs are currently known, and they are introduced later in the paper), is essential as well as multi-wavelength follow-up studies such as those by \citet{Norris2022},  \citet{Rupke2023} and \citet{Coil2024}. \\
 
In clusters with double-relics the face-on orientation minimizes projection effects and provides better constraints on shock model parameters (e.g., spectral gradients, injection spectral indices, shock Mach numbers) and the physical properties of the underlying merging system (e.g., mass, mass ratio). The main open questions involving radio relics concern their efficiency in particle acceleration, their relation with shock waves and sources of cosmic-ray electrons (such as AGN), the magnetic field strength and orientation, as well as the merger state of the host clusters. For numerical simulations of double relics, including at different sky projections, see \cite{vanWeeren11_sim}, \cite{Bonafede12}, \cite{sk13} and \cite{Wittor17b, Wittor21}. In the southern sky, there are several double-relic clusters known, including Abell~2345 \citep{Bonafede2009}, Abell~3376 \citep{Bagchi06,Chibueze2023}, MACS~J0025.4-1222 \citep{Riseley2017}, El Gordo \citep{Lindner2014}, PLCK G287.0+32.9 \citep{Bagchi2011}, RXC~J1314.4--2515 \citep{Feretti2005}, Abell~3365 \citep{Stuardi2022,Urdampilleta2021}, and Abell~3667 \citep{Roettgering1997,deGasperin2022}. The MeerKAT Galaxy Cluster Legacy Survey \citep[MGCLS,][]{Knowles2022}, which targeted 115 galaxy clusters for $\sim$6--10~h each, contains seven systems with double relics, all but one with radio halos. The five new systems are MCXC J0352.4--7401, Abell~521, MCXC J0516.6--5430, MCXC J0232.2--4420, and RXC J2351.0--1954. The LOFAR Two-meter Sky Survey data release two (LOTSS-DR2) was used for a study of diffuse radio emission (halos and relics) in 309 PSZ2 clusters plus a comparison to X-ray data from XMM and Chandra \citep{Botteon2022,Jones2023}; our PSZ2 cluster is not part of this study. Among these are six double relics (PSZ2 G071.21+28.86, PSZ2 G099.48+55.60, PSZ2 G113.91--37.01, PSZ2 G165.46+66.15, PSZ2 G181.06+48.47, and PSZ2 G205.90+73.76). Further PSZ2 clusters were studied by \citet{Duchesne2024} using ASKAP radio continuum data from the Evolutionary Map of the Universe \citep[EMU,][]{EMUPS} pilot survey. In addition there have been searches for diffuse radio sources in non-PSZ clusters, e.g. \citep{Hoang2022} find at least two symmetric (i.e. on opposite sides of the cluster) double relics (in Abell~373 and Abell~1889), \citep{Duchesne2021} find a symmetric double relic in SPT-CL J2032-5627), and \citep{Knowles2021} study an asymmetric double relic in ACT-CL J0159.0--3413. \\

In this paper we present two discoveries: (1) the large double relic around the galaxy cluster PSZ2~G277.93+12.34, and (2) ORC J1027--4422, both found in the same MeerKAT 1.3 GHz image. These are presented together as their likely formation mechanisms both involve merger shocks. \\

We use eROSITA data to search for extended X-ray emission in the PSZ2~G277.93+12.34 cluster in order to characterize the gaseous cluster properties in more detail. The cluster was marginally detected via the Sunyaev-Zeldovich (SZ) effect at a signal to noise of 5.1 by the \citet{Planck2016}. It has an integrated flux of $Y_{5R500} = (1.87 \pm 0.90) \times 10^{-3}$ arcmin$^2$ and is one of over 1000 SZ sources detected in the Planck full-mission data. For a cluster redshift of $z \approx 0.158$ (see Section~3.1), which corresponds to a luminosity distance of $D_{\rm L}$ = 755~Mpc, we derive an SZ mass of $M_{500}$ = ($3.6 \pm 0.6$) $\times$ 10$^{14}$\Msun\ \citep{Arnaud2010}. A similar double radio relic, detected in the merging galaxy cluster PSZ2~G096.88+24.18 ($z \approx 0.3$), was recently studied in detail by \citet{Jones2021} using VLA 1.5~GHz, LOFAR 140~MHz and Chandra X-ray data. They found that the two diametrically opposed radio relics of PSZ2~G096.88+24.18 with a separation of $\sim$2~Mpc are approximately equidistant from the cluster center. This sets a good example of a single object, combined radio / X-ray study similar to our study of PSZ2~G277.93+12.34 using MeerKAT, ASKAP and Murchison Widefield Array (MWA) radio data as well as eROSITA X-ray data. \\
  
We analyse the properties of ORC~J1027--4422 and compare its morphology to that of cluster relics. One possible explanation for their formation is discussed by \cite{Dolag2023} who find ORC-like galaxy merger shocks in their high-resolution cosmological simulations, similar but smaller than typical cluster merger shocks. \\ 

The paper is organised as follows: in Section~2 we describe the MeerKAT, ASKAP and eROSITA data acquistion and processing. Our results on the PSZ2~G277.93+12.34 cluster are given in Section~3, followed by the analysis of ORC~J1017--4422 in Section~4. Discussion and conclusions follow in Sections~5 and 6, respectively. We adopt a $\Lambda$CDM cosmology with $H_{\rm 0}$ = 70\kms\,Mpc$^{-1}$, $\Omega_{\rm M}$ = 0.3 and $\Omega_{\rm \Lambda}$ = 0.7. 
% https://cosmocalc.icrar.org/

\begin{figure*} % Figure 1
\centering
  \includegraphics[width=15cm]{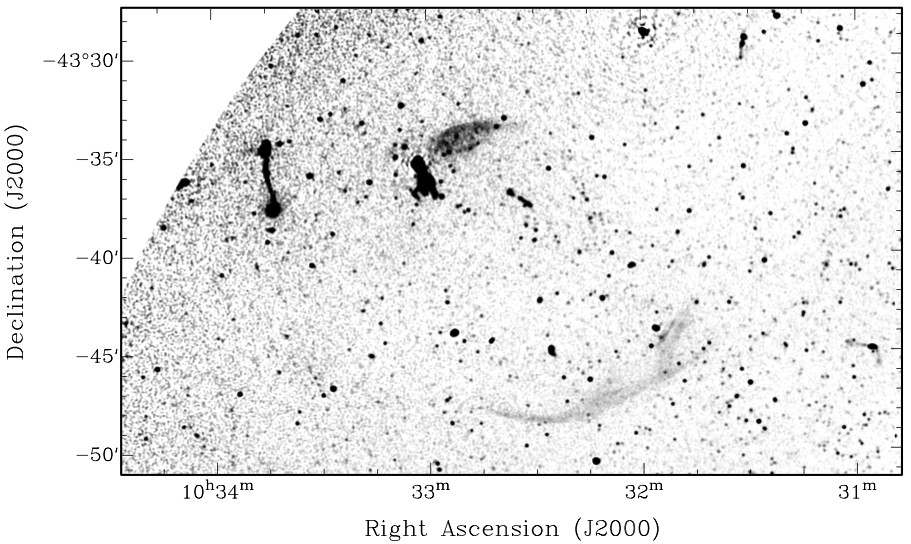}
\caption{MeerKAT 1.3~GHz wide-band radio continuum image of a newly discovered double relic associated with the galaxy cluster PSZ2~G277.93+12.34. The image has a resolution of 7.7\arcsec\ and is primary beam corrected. The pointing center lies $\sim$43\arcmin\ south-west of the cluster centre.}
\label{fig:fig1}
\end{figure*}

\begin{figure*} % Figure 2
\centering
  \includegraphics[width=15cm]{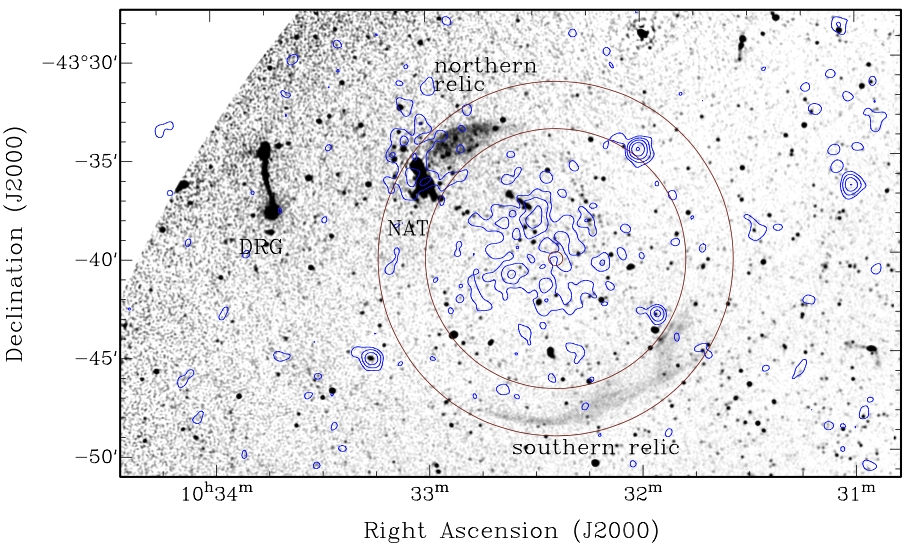}
\caption{As Fig.~\ref{fig:fig1}, but now overlaid with eRASS:3 X-ray (0.3 -- 2 keV) contours in blue. The eROSITA image was smoothed with a 30\arcsec\ Gaussian; the contour levels are 0.001, 0.002, 0.004, 0.008 and 0.016 cts\,s$^{-1}$. The overlaid circles indicate the approximate width and centre of the double relic. Both relics, the narrow-angle tail (NAT) radio galaxy PMN J1033--4335 and the double-lobe radio galaxy (DRG) to the east are labelled.}
% The overlaid circles have diameters of 14.4 arcmin and 17.2 arcmin, centred on 158.1 deg, --43.67 deg (by eye estimate).
\label{fig:fig2}
\end{figure*}

\begin{figure*} % Figure 3
\centering
\includegraphics[width=14cm]{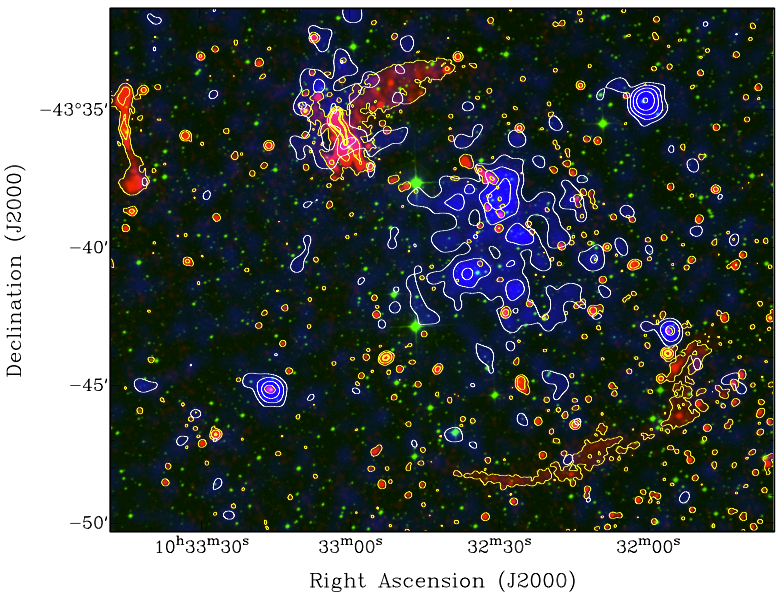}
\caption{RGB colour image of the double relic cluster PSZ2~G277.93+12.34. The MeerKAT 1.3 GHz radio continuum emission is shown in red plus yellow contours (0.01, 0.1 and 1 mJy\,beam$^{-1}$), the eRASS:3 X-ray emission in blue plus white contours (same levels are as in Fig.~\ref{fig:fig2}), and the DSS2 optical $R$-band image in green. For display purposes we show the radio emission without primary beam correction.}
\label{fig:relic-RGB}
\end{figure*}

\section{Observations and Data Processing}

\subsection{MeerKAT}
\label{sec:obs}

MeerKAT is a powerful SKA precursor radio interferometer located in the Karoo desert of South Africa. It consists of 64 $\times$ 13.5~m antennas, with baselines up to 8~km \citep{Jonas2009,Jonas2016,Mauch2020}. Of these, 48 antennas are located in the inner core (within a 1-km radius) with the shortest baseline being 29~m. Our field, targeting the nearby NGC~3263 galaxy group (Koribalski et al. 2024, in prep.), was observed on the 27th of May 2021 with 59 out of 64 MeerKAT antennas for 4.64~h and on the 5th of Jan 2022 with 62 antennas for 4.86~h (project ID: SCI-20210212-BK-01). The observations were split into two $\sim$5~h blocks to allow for easy scheduling, with the blocks covering complementary LST ranges to ensure full $uv$-coverage. The pointing centre is at $\alpha,\delta$ (J2000) = $10^{\rm h}\,29^{\rm m}\,13^{\rm s}$, --44\degr\,07\arcmin\,20\arcsec, and the central frequency is 1284~MHz. In the high-resolution spectral line mode the bandwidth of 856~MHz is divided into 32k channels. The resulting frequency range is 856 to 1712 MHz, overlapping at the low frequency end with the ASKAP data (see Section~2.2). The fully calibrated continuum data products provided by SARAO include a wide-band radio continuum image, an in-band spectral index map, and an image cube consisting of roughly constant fractional bandwidth sub-band images, i.e. not equally spaced in frequency \citep{Cotton2008,Cotton2019}\footnote{See https://www.cv.nrao.edu/$\sim$bcotton/ObitDoc/MFImage.pdf}; these products are not primary-beam corrected. A primary beam corrected wide-band radio continuum image is also provided. About 50\% of the MeerKAT band was flagged due to the presence of radio frequency interference (RFI); see, for example, \cite{Heywood2022}. The primary flux and bandpass calibrators were PKS J0408--6545 (model intensity: 15.713 Jy) and PKS J1939--6342 (model intensity: 15.006 Jy). The source PKS J1120--2508 (measured intensity: 1.746 Jy) was used as the secondary calibrator. The target field and the secondary calibrator were observed alternately for 36 and 2 minutes. \\

To create our final radio continuum images we used two procedures. Initially, we combined the two fully calibrated Stokes~$I$ MeerKAT images and, separately, the 10-channel cubes obtained from the SARAO archive, made using the standard MeerKAT continuum pipeline. For the latter, a frequency-dependent $uv$-taper was used to ensure the resolution is the same in all 10 sub-band channels \citep[see][and references therein]{Mauch2020}. The resulting cube is used to explore the in-band spectral indices of the discovered radio sources. The full-bandwidth Stokes~$I$ radio continuum images are made using multi-frequency synthesis, which ensures excellent $uv$-coverage and sensitivity to extended structures. We use tasks in the \texttt{miriad} software package for data combination and analysis. The measured rms near the cluster relics is 3.2 $\mu$Jy\,beam$^{-1}$ (Day~1, beam: $7.63\arcsec \times 6.68\arcsec$) and 3.5 $\mu$Jy\,beam$^{-1}$ (Day~2, beam: $7.45\arcsec \times 7.14\arcsec$). We combine the two images after convolving both to a common resolution of 8\arcsec. This results in an improved rms of 2.4~$\mu$Jy\,beam$^{-1}$. All image analysis is done using the {\sc miriad} package \citep{Sault1995}. Fluxes are measured using the task {\sc cgcurs} with appropriate regions.  \\

In our second approach, to improve the quality of the Stokes~$I$ wide-band image and extend the primary beam correction to the 5\% level, we downloaded the two calibrated data sets and jointly imaged them. The method and scripts used to process MeerKAT continuum observations are provided online \citep{Heywood2020}\footnote{https://github.com/IanHeywood/oxkat}. Since the data were transferred from the archive pre-calibrated and flagged, we performed a single round of phase and delay direction independent calibration using {\sc wsclean} \citep{offringa2014} for imaging and {\sc cubical} \citep{kenyon2018} for calibration. Direction dependent corrections were derived using {\sc killms} \citep{smirnov2015,tasse2023a} and applied using the {\sc ddfacet} imaging package \citep{tasse2018,tasse2023b}. With the exception of the primary beam cut, all parameters for calibration and imaging we the same as those used by \citet{Heywood2022}. We achieved a synthesized beam of 7.7\arcsec\ and a much improved rms of $\sim$1.5 $\mu$Jy\,beam$^{-1}$ around the field centre. The rms noise between the two relics is $\sim$6.5 $\mu$Jy\,beam$^{-1}$ and near ORC~J1027--4422 it is $\sim$4.0 $\mu$Jy\,beam$^{-1}$, higher than around the image centre due to the primary beam correction. All MeerKAT radio continuum images shown in this paper are from this approach. \\

The MeerKAT primary beam full width half maximum (FWHM) is approximately $67.9\arcmin \times 65.3\arcmin$ at 1.3~GHz; it varies from $\sim$100\arcmin\ at the low frequency end of the L-band to $\sim$55\arcmin\ at its high frequency end \citep{Mauch2020, Heywood2022, deVilliers2022}. The northern relic and PMN J1033--4335 are located $\sim$52\arcmin\ from the pointing centre, i.e. 1.5$\times$ the half power beam width (HPBW) at 1.3~GHz, while the southern relic lies roughly at the 1.3~GHz HPBW, and ORC~J1027--4422 resides $\sim$26.4\arcmin\ from the pointing centre ($\sim$0.8$\times$ HPBW at 1.3~GHz). This means that our flux estimates at the high-frequency end of the band, where the primary beam correction is large and our sensitivity to extended structures is somewhat diminished, have large uncertainties (for details see Section~3.1.1). Consequently, reliable in-band spectral index and polarisation analysis of the above sources is not feasible. 

\subsection{ASKAP}
We supplement our deep MeerKAT 0.9--1.7~GHz data with Stokes~$I$ radio continuum images at $\sim$0.9~GHz from the Australian SKA Pathfinder \cite[ASKAP,][]{Johnston2008, Hotan2021, Koribalski2022}, obtained in March 2020 as part of the Rapid ASKAP Continuum Survey \cite[RACS,][]{McConnell2020}. The first release of RACS-low covers the sky south of $\delta \approx +41$\degr\ at a central frequency of 887.5~MHz with 288~MHz bandwidth, overlapping with the low frequency end of the MeerKAT data. We use the mosaicked RACS-low images at 25\arcsec\ resolution and measure an rms of 0.3~mJy\,beam$^{-1}$ near the double relic and near ORC~J1027--4422. We use the RACS-low data to compare with and confirm the primary-beam corrected channel~1 (888~MHz) MeerKAT flux densities of the northern cluster relic and ORC~J1027--4422.

% Radio fluxes of PMN J1033-4335
% 4850 MHz: 0.080 +- 0.010 Jy
% 1300 MHz: 0.234 +- 0.003 Jy 
%  888 MHz: 0.290 +- 0.005 Jy
%  888 MHz: 0.303 +- 0.005 Jy (Hale)
%  843 MHz: 0.322 +- 0.012 Jy
%  200 MHz: 0.860 +- 0.031 Jy
%  150 MHz: 0.883 +- 0.089 Jy

Recently released ASKAP data from RACS-mid \citep{Duchesne2023} at 1367.5~MHz (bandwidth 144~MHz) are used for further flux comparison. 
% At the cluster position their angular resolution is $\sim$9\arcsec, and we measure an rms of 0.24~mJy\,beam$^{-1}$.

\subsection{eROSITA}
eROSITA is a new X-ray telescope onboard the Spectrum-Roentgen-Gamma \citep[SRG,][]{Sunyaev2021} space observatory which was launched in mid 2019. It consists of seven mirrors and aims to conduct an all-sky survey program (the eROSITA All Sky Survey: eRASS) every half year. The eRASS data products are offered as 4,700 overlapping sky tiles, each of which has a size of $3.6\degr \times 3.6\degr$. The eROSITA telescope mirrors provide an average point spread function (PSF) of around 26\arcsec\ half power diameter (HPD) \citep{Predehl21}.

\begin{table} % Table 1
\centering
\caption{Positions of the eROSITA sky tiles used for the X-ray emission analysis of the PSZ2~G277.93+12.34 galaxy cluster and ORC~J1027--4422.}
\begin{tabular}{ccc}
\hline
Tile ID & RA (J2000) & Dec (J2000) \\
\hline
\hline
sm159135 & 159.3103 & --45.0091 \\
sm160132 & 160.2198 & --42.0083 \\
sm155135 & 155.1724 & --45.0091 \\
sm156132 & 156.2637 & --42.0083 \\
\hline
\end{tabular}
\label{tab:tab1}
\end{table}

The PSZ2 G277.93+12.34 galaxy cluster is mainly located in the eROSITA sky tile sm159135. To account for possible emission from the cluster surroundings, we also made use of three adjacent sky tiles. All four sky tiles and their center coordinates are listed in Table~\ref{tab:tab1}. We combined data from three eROSITA sky surveys (eRASS1--3, the combination of all three is called eRASS:3), centered on the PSZ2 G277.93+12.34 cluster and restricted to $\sim$1~degree radius from the center. The internal data processing version c946 was used together with the extended Science Analysis Software \citep[eSASS,][]{Brunner2022} version 211214. For details of the data reduction and image correction steps we refer the reader to Section~2.1 of \cite{Reiprich_2021}.

\subsubsection{Data Reduction and Image Creation}

The eROSITA data reduction began with generating the clean event files and images using the \texttt{evtool} tasks. We specified \texttt{flag=0xc00fff30}, which removes bad pixels and the strongly vignetted corners of the square CCDs, and \texttt{pattern=15} to include all patterns (single, double, triple, and quadruple). To analyze the hot gas emission, we focus the imaging analysis on the energy band of 0.3--2.0~keV (hereafter, soft band). The lower energy limit used for the telescope modules (TMs) with on-chip filter (TM1--4, 6; the combination of these TMs is referred to as TM8) was set to 0.3 keV, while for the TMs without on-chip filter (TM5 and 7; the combination is referred to as TM9) due to the optical light leak contamination \citep{Predehl21}, the lower energy limit was set to 0.8~keV. The next step was to subtract the particle-induced background (PIB) from the image. We modeled the PIB for each TM in each observation based on the results of the eROSITA Filter-Wheel-Closed (FWC) observation data. As the temporal variability of the PIB spectral shape appears to be very small, the count rate in a hard band should be strongly dominated by PIB events. Hence, we used the counts in the 6--9~keV band (hereafter, hard band) from our observations as the total PIB counts. By multiplying these hard band counts with the soft-to-hard band counts ratio from the FWC data, we obtain the PIB count estimates in the soft band. Afterwards, the soft band PIB counts were spatially distributed by multiplying it to the non-vignetted exposure map, which was normalized to unity by dividing each pixel by the sum of all pixel values.

The final cleaned and exposure corrected count rate image (see Figs.~2 \& 3) was obtained by dividing all observation combined PIB-subtracted count image by the final combined and corrected TM0 exposure map. We note that the count rates of the final combined image correspond to an effective area given by one TM with an on-chip filter in the energy band 0.3--2.0~keV.

\subsubsection{Spectral Analysis}
We performed spectral analysis with eRASS:3 data to obtain estimates of the ICM properties of the main (m) structure of the PSZ2 G277.93+12.34 galaxy cluster and the northern (n) structure. All eROSITA spectra from the seven TMs are extracted using the eSASS task \texttt{srctool}. For the main and the northern structure, the X-ray spectra are extracted from circles with radii of 5.7\arcmin\ (0.93 Mpc) and 4.6\arcmin\ (0.73 Mpc), respectively, centered at the X-ray centre positions given in Table~\ref{tab:tab2}. The spectral fitting was realized with \texttt{XSPEC} \citep{Arnaud1996} version 12.12.0. The model for the spectral fitting \citep[for other eROSITA spectral fitting examples see, e.g.,][]{Ghirardini2021, Iljenkarevic2022, Veronica_2022} includes the cosmic X-ray fore/background (CXB) and the source emission and describes as follows 
\begin{equation}
\begin{split}
\mathtt{Model =} &\quad\mathtt{constant\times(apec_1 + TBabs\times(apec_2 +}\\
&\quad\mathtt{powerlaw) + TBabs\times apec_3}).\\
\end{split}
\label{eq:spectral_model}
\end{equation}

\begin{figure} % Figure 4
\centering
\includegraphics[width=8cm]{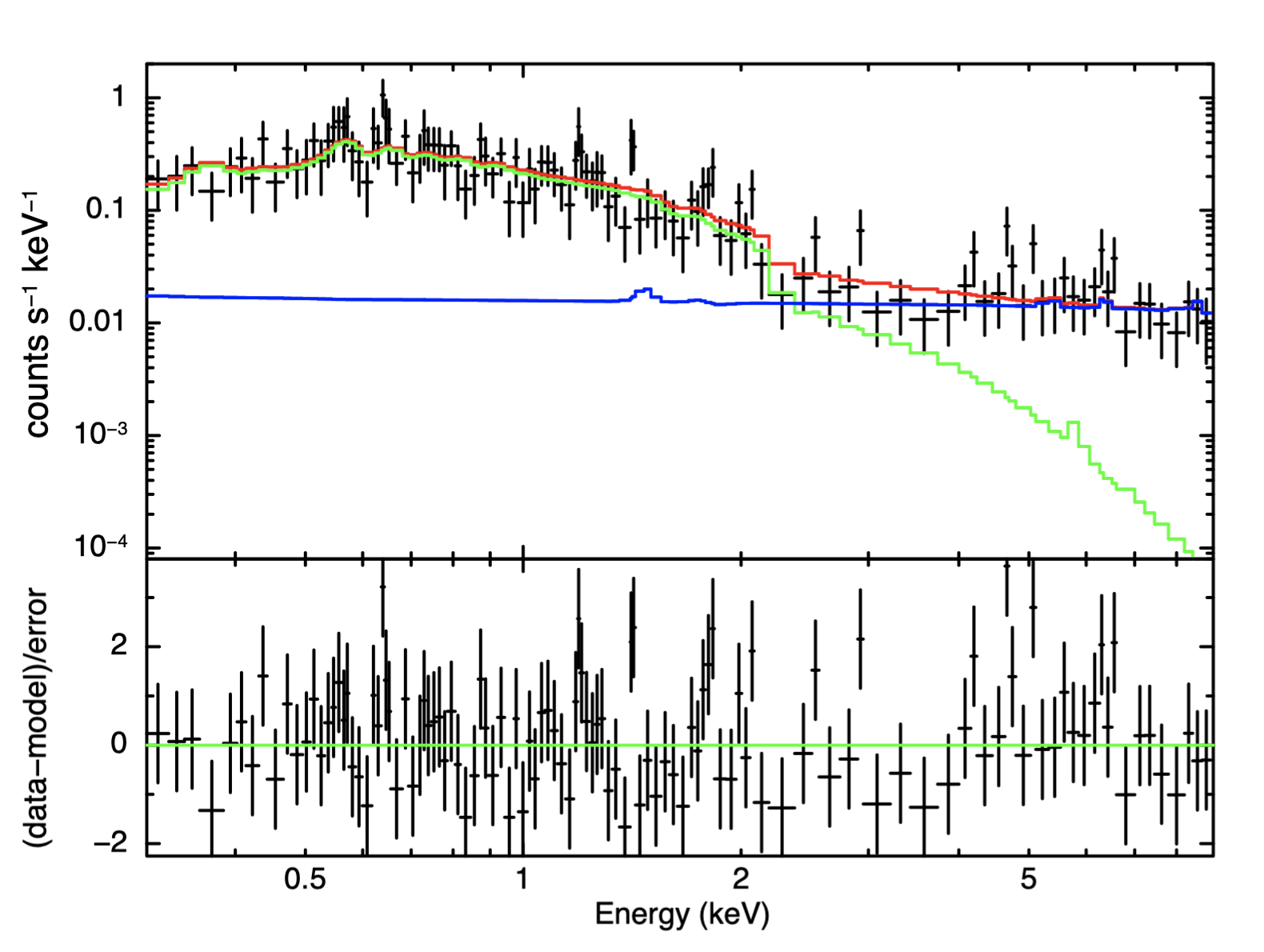}
\includegraphics[width=8cm]{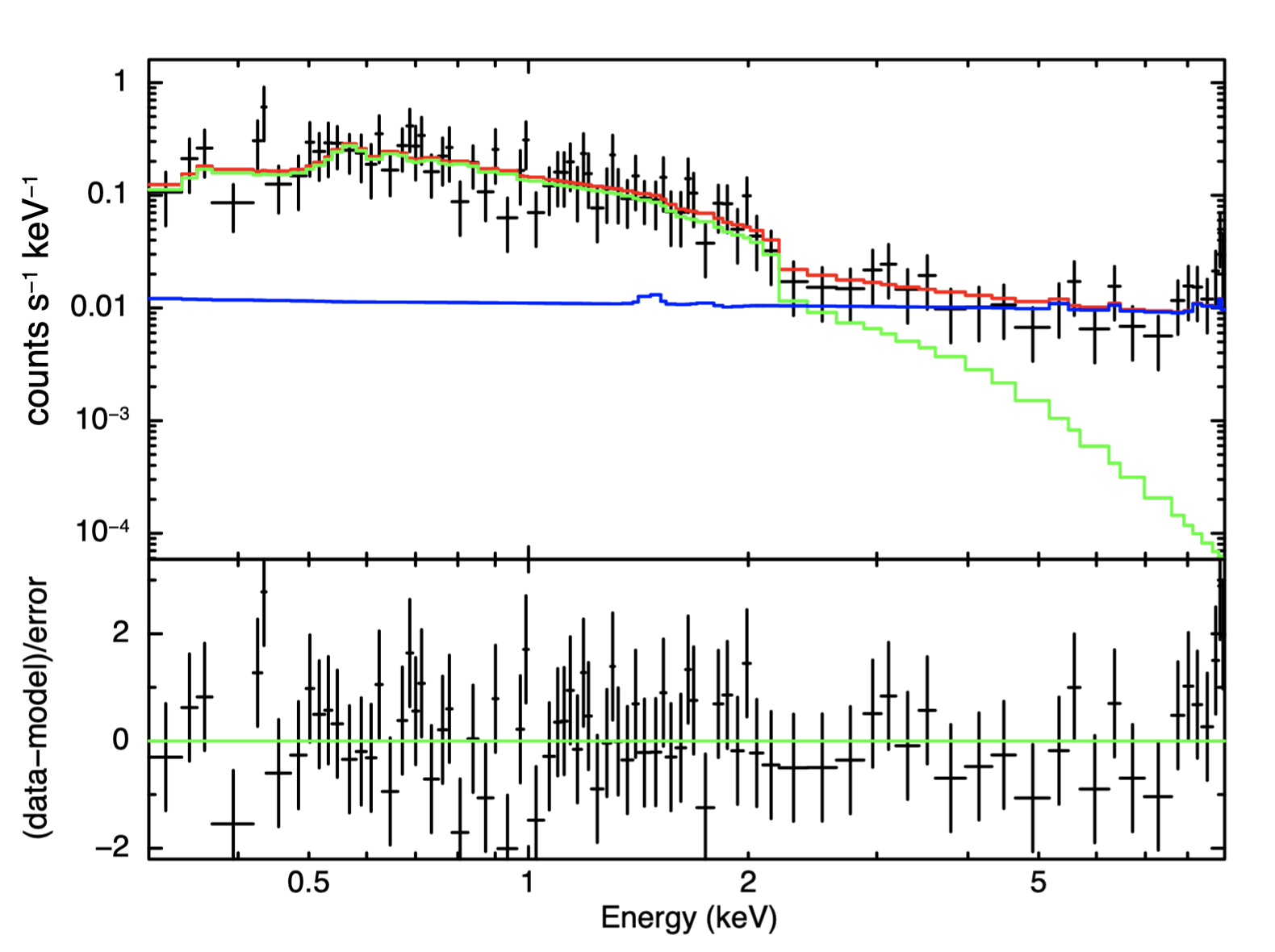}
\caption{eRASS:3 integrated X-ray spectra for the main cluster structure (top) and the northern structure (bottom) in the energy band of $0.3-9.0$~keV. The data points were extracted from the combined telescope modules with the on-chip filter (TM8, see Section~2.3). In the upper part of each plot we show the eRASS:3 data points (black crosses), the total model (red line), the source plus sky background emission (green line), and the instrumental background (blue line), while the residual emission spectrum in terms of sigma deviation is shown in the bottom part.}
\label{fig:spectra}
\end{figure}

The first terms of the equation depict the CXB components scaled to the areas of the source regions (\texttt{constant} [arcmin$^2$]). The absorption along the line of sight is represented by \texttt{TBabs} \citep{Wilms_2000}. The adopted $N_{\rm HI}$ values used in this work are from the column density map of the neutral atomic hydrogen (\HI) by the \cite{HI4PI2016}. The thermal emission from the Local Hot Bubble (LHB) and the Milky Way Halo (MWH) are represented by \texttt{apec$\mathtt{_1}$} and \texttt{apec$\mathtt{_2}$}, where we fixed their temperature at $k_\mathrm{B}T$ = 0.1 and 0.25 keV, respectively. The absorbed thermal emission of the source spectra is represented by \texttt{TBabs $\times$ apec$\mathtt{_3}$}. The cosmic X-ray background from the unresolved sources \citep[e.g.,][]{Luo_2017} is characterized by a \texttt{powerlaw} with photon index of 1.46. Additionally, the results of the eROSITA EDR FWC\footnote{\href{https://erosita.mpe.mpg.de/edr/eROSITAObservations/EDRFWC/}{https://erosita.mpe.mpg.de/edr/eROSITAObservations/EDRFWC/}} data analysis are used for modeling the instrumental background. The instrumental background model includes two power-laws for the increase of the background due to the detector noise in the low energy range, a combination of a power-law and an exponential cut-off for the signal above $\sim$1~keV, and 14~Gaussian lines for the X-ray fluorescence lines from the interaction between cosmic particles and the detector components. We refer to \cite{Freyberg2022} for an overview of the eROSITA in-flight background. The normalizations of the CXB components and the instrumental background components of the source spectra are left to vary throughout the fit. The fit was performed in the energy band of $0.3-9.0$ keV for the TMs with the on-chip filter and $0.8-9.0$ keV for the TMs without the on-chip filter. We adopted the C-statistics \citep{Cash_1979} and the Solar abundance table from \cite{Asplund_2009}. The properties (centre position, radius and $N_{\rm HI}$) of the two X-ray emission structures are tabulated in Table~\ref{tab:tab2} and example spectra are shown in Fig.~\ref{fig:spectra}.

\begin{table} % Table 2
\centering
\caption{Properties of the eRASS:3 X-ray emission structures associated with the galaxy cluster PSZ2~G277.93+12.34. We give the centre positions and radii of the circular areas from which the X-ray spectra are extracted as well as the respective \HI\ column densities, $N_{\rm HI}$.} 
\begin{tabular}{cccc}
\hline
structure & RA, Dec (J2000) & radius & $N_{\rm HI}$ \\
 & [degr], [degr] & [arcmin] & [cm$^{-2}$] \\
\hline
\hline
main (m)  & 158.1258, --43.6502 & 5.7 & $7.8 \times 10^{20}$ \\
north (n) & 158.2644, --43.5808 & 4.6 & $7.6 \times 10^{20}$ \\
\hline
\end{tabular}
\label{tab:tab2}
\end{table}
% RA,DEC(J2000) 
% main: 10:32:30.192, -43:39:00.72
% north: 10:33:03.456, -43:34:50.88

\begin{table*} % Table 3
\centering
\begin{tabular}{ccccc}
\hline
  Source name & redshift & X-ray & Notes & separation \\
\hline
\hline
(1) main cluster structure & 0.158 & radius 5\farcm7 & (see Fig.~\ref{fig:psz2-center}) \\
WISEA J103230.00--433815.4 & 0.1579 (p) & extended & likely BCG & --- \\
WISEA J103236.75--433625.6 & 0.2060 (p) & & & 2.2$'$ \\  
WISEA J103228.20--433530.0 & 0.1839 (p) & & & 2.8$'$ \\ 
WISEA J103243.59--434042.5 & 0.2034 (p) & & & 3.5$'$ \\
WISEA J103302.46--433508.5 & 0.1513 (p), 0.1561 (p) & & NAT host galaxy & 6.6$'$ \\
WISEA J103225.84--434454.1 & 0.1242 (p) & & & 6.7$'$ \\
\hline
(2) northern cluster structure & 0.153 & radius 4\farcm6 & (see Fig.~\ref{fig:nat}) \\
WISEA J103302.46--433508.5 & 0.1513 (p), 0.1561 (p) & & NAT host galaxy & --- \\
WISEA J103305.55--433412.1 & $0.15454 \pm 0.00015$ (s) & & LEDA 547219 & 1.1$'$ \\
WISEA J103307.98--433625.2 & $0.15776 \pm 0.00015$ (s) & & LEDA 546771 & 1.6$'$ \\
2MASS J10334526--4335420 & 0.1584 (g) & & DRG & 7.8$'$ \\
\hline
\end{tabular}
\caption{Galaxies likely associated with the main and northern structures of the PSZ2 G277.93+12.34 cluster. In Col.~2 we list photometric (p) and spectroscopic (s) redshifts from \citet{Bilicki2014,Bilicki2016} and \citet{Jones2009}, respectively. (g) denotes redshifts from the \citet{Gaia2022}.}
\label{tab:tab3}
\end{table*}

\section{Cluster results}

The PSZ2~G277.93+12.34 galaxy cluster members are currently not well defined as redshifts are available for only a few galaxies in the area. We adopt $z \approx 0.158$ as the cluster redshift based on the photometric redshift of the likely brightest cluster galaxy, WISEA J103230.00--433815.4. Galaxies in the vicinity indicate that the cluster redshift could be somewhat higher (see Table~\ref{tab:tab3}).

The host galaxy of the nearby NAT radio galaxy, PMN J1033-4335, located $\sim$7\arcmin\ north-east of the PSZ2 cluster position, is WISEA~J103302.46--433508.5 (2MASX J10330244--4335085). Two photometric redshifts are reported, \zph\ $\approx$ 0.1513 \citep{Bilicki2014} and \zph\ $\approx$ 0.1561 \citep{Bilicki2016}, which gives an average value of \zph\ $\approx$ 0.153, adopted here. Two galaxies in the NAT's vicinity, also listed in Table~\ref{tab:tab3}, have similar spectroscopic redshifts.

\subsection{MeerKAT radio continuum emission}
\begin{figure} % Figure 5
\centering
   \includegraphics[width=8.5cm]{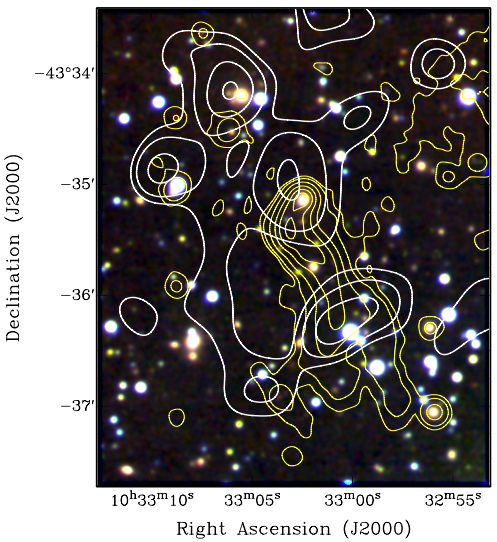}
\caption{MeerKAT 1.3 GHz radio continuum emission of the NAT radio galaxy PMN J1033--4335 (yellow contours, resolution 7.7\arcsec, primary beam corrected) and eROSITA X-ray emission (white contours; smoothed with a 30\arcsec\ Gaussian) overlaid onto an RGB colour image consisting of VHS $K_{\rm s}$-band (red), VHS $J$-band (green) and DSS2 $R$-band (blue). The NAT's host galaxy is WISEA J103302.46--433508.5 at \zph\ $\approx$ 0.153 \citep{Bilicki2014,Bilicki2016}. The contour levels are 0.06, 0.2, 1, 4, 10 and 20 mJy\,beam$^{-1}$ (radio) and 0.0012, 0.0015, 0.002 and 0.0025 cts\,s$^{-1}$ (X-ray). To allow comparison between the radio and X-ray morphologies, only the bright emission regions are shown here.}
\label{fig:nat}
\end{figure}

\begin{figure*} % Figure 6
\centering
\includegraphics[width=14cm]{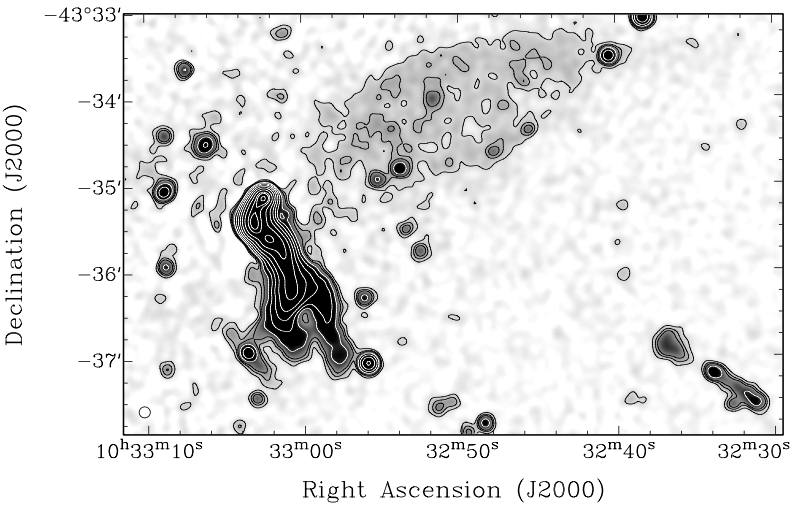}
\caption{MeerKAT 1.3 GHz radio continuum emission (primary beam corrected) of the NAT radio galaxy and the northern cluster relic. The contour levels are 0.04, 0.08, 0.12 (black) and 0.25, 0.5, 1, 2, 4, 8, 15, 20 and 30 (white) mJy\,beam$^{-1}$. The synthesized beam (7.7\arcsec) is shown in the bottom left corner.}
\label{fig:Nrelic}
\end{figure*}

\begin{figure*} % Figure 7
\centering
\includegraphics[width=14cm]{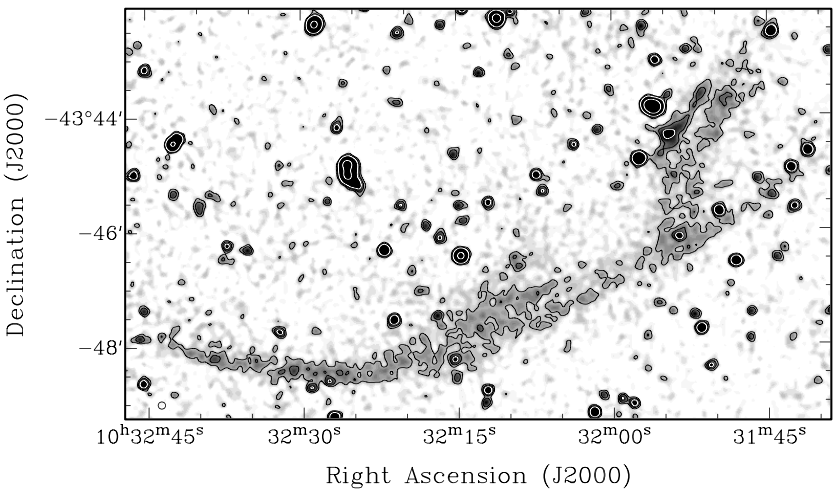}
\caption{MeerKAT 1.3 GHz radio continuum emission (primary beam corrected) of the southern relic, which extends over at least 10\arcmin. The contour levels are 0.015, 0.03 (black), 0.06, and 0.6 (white) mJy\,beam$^{-1}$. The synthesized beam (7.7\arcsec) is shown in the bottom left corner.}
\label{fig:Srelic}
\end{figure*}

Figures~\ref{fig:fig1}--\ref{fig:relic-RGB} show the MeerKAT 1.3 GHz radio continuum images of the PSZ2~G277.93+12.34 galaxy cluster and surroundings. We found two large radio relics, located NE and SW of the cluster centre, forming a double radio relic with an angular separation of $\sim$16~arcmin, as well as a prominent narrow-angle tail (NAT) radio galaxy, PMN J1033--4335, just east of the northern relic. The two arc-shaped relics, which consist of diffuse, non-thermal synchrotron emission (see below), form part of a circle as indicated in Fig.~\ref{fig:fig2}, likely tracing shocks induced by a face-on cluster merger. The circle centre (158.1\degr, --43.67\degr) approximately agrees with that of the X-ray emission from the main cluster structure (see Table~\ref{tab:tab2}). For a cluster redshift of $z \approx 0.158$ we estimate a double relic separation of $\sim$2.6~Mpc. Figures~\ref{fig:nat}--\ref{fig:psz2-center} show close-ups of the NAT radio galaxy, the two relics, and the central cluster area.

\subsubsection{The NAT radio galaxy PMN J1033--4335}

The most prominent radio galaxy in the cluster field is PMN J1033--4335, situated just east of the northern cluster relic (see Fig.~\ref{fig:fig1}--\ref{fig:relic-RGB}). It is very bright and well-resolved by both ASKAP and MeerKAT. We find its radio morphology that of a head-tail or narrow-angle tail (NAT) radio galaxy (see Fig.~\ref{fig:nat}). Radio lobes trailing south-wards from the radio core (head) suggest the galaxy is rapidly moving away from the cluster centre. The projected angular size of the NAT's radio lobes is $\sim$2\arcmin\ or $\sim$320~kpc at the assumed cluster redshift. The NAT radio core and the northern relic have the same projected distance from the cluster centre. It seems plausible that the NAT radio galaxy has fallen through the cluster core and is moving outwards with high speed, possibly overtaking the northern relic. While the NAT may be embedded in the northern relic, it is more likely seen in projection against the relic's eastern part. We measure a total flux density of $234 \pm 3$~mJy in our primary-beam corrected MeerKAT 1.3~GHz wide-band image. 

Using the RACS-low 888~MHz image we measure a flux density of $290 \pm 5$~mJy for the NAT radio galaxy, in agreement with the SUMSS 843~MHz flux density of $321.7 \pm 11.6$ mJy and the catalogued RACS-low flux density of $303.3 \pm 4.6$~mJy \citep{Hale2021}. PMN J1033--4335 is also detected in the 150~MHz TIFR GMRT Sky Survey (TGSS; resolution $\sim$25\arcsec, rms $\sim$3.5 mJy\,beam$^{-1}$) with a flux density of $883 \pm 89$ mJy \citep{TGSSADR2016,Intema2017}. Using low-frequency images from the Galactic and Extragalactic All Sky MWA \citep[GLEAM,][]{Hurley-Walker2017} survey, which has a resolution of $\sim$2\arcmin\ at 200~MHz, we estimate integrated 140 -- 170 MHz and 170 -- 230~MHz flux densities of $1025 \pm 36$~mJy and $860 \pm 31$~mJy, respectively. Using the GLEAM ($860 \pm 31$ mJy at 200~MHz), the RACS-low ($290 \pm 5$ mJy at 888~MHz) and the MeerKAT ($234 \pm 3$~mJy at 1.3~GHz) measurements, we derive a spectral index of $\alpha \approx -0.71 \pm 0.04$, where $S_{\nu} \propto \nu^\alpha$.

\subsubsection{Northern Relic}

The northern relic extends over at least 4\arcmin\ (660~kpc) and has a convex outer edge with respect to the cluster location, as expected. It is shorter and wider ($\sim$90\arcsec) than the southern relic by at least a factor two. The NAT radio galaxy overlaps (likely in projection) with the eastern side of the northern relic. We measure a flux density of $15 \pm 3$~mJy in the primary beam corrected MeerKAT 1.3~GHz wide-band data. This value is somewhat uncertain because of the relic's low-surface brightness and large distance from the pointing centre ($51.4 \pm 0.8$\arcmin\ or $\sim$1.5 $\times$ the HPBW at 1.3~GHz where the sensitivity has dropped to $\sim$19\% $\pm$ 1\%). The corresponding radio power is $P_{1300} \sim 10^{24}$~W\,Hz$^{-1}$ at the adopted cluster distance.

% Radio power calculation:
% P_radio [W/Hz] = 4 pi D_L^2 x S_radio
%  without k-correction (1+z)^(alpha-1)

Using the RACS-low 888~MHz and GLEAM 200~MHz images we obtain flux densities of $13 \pm 1$~mJy and $43 \pm 7$~mJy for the northern relic, respectively, corresponding to radio powers of $P_{888} \sim 0.9 \times 10^{24}$~W\,Hz$^{-1}$ and $P_{200} \sim 2.9 \times 10^{24}$~W\,Hz$^{-1}$. The latter follows the trend shown by \citet[][their Fig.~6]{Jones2023}. We estimate a spectral index of $\alpha \approx -0.8 \pm 0.1$ between RACS-low and GLEAM. While we detect the northern relic in the low-resolution GLEAM images, it is not seen in the shallow TGSS 150~MHz image due to the relic's low surface brightness.
% alpha = log(S1/S2) / log(nu1/nu2)
% --- Northern relic ---
% = log(13/43) / log(888/200) = -0.8

\citet[][see their Tables~2 \& 3]{Jones2023} find a radio relic power -- cluster mass relation of
\begin{equation}
 \log_{10} (P_{150}) = 5.2 \cdot \log_{10} (M_{500}/M_{\odot}) - 51.3~,
 \end{equation}
and a radio relic power -- LLS relation of 
\begin{equation}
    \log_{10} (P_{150}) = 7.6 \cdot \log_{10} {\rm (LLS/kpc)} + 2.15~,
 \end{equation}
based on their orthogonal distance regressions. Using $P_{150} = 2.9 \times 10^{24}$~W\,Hz$^{-1}$, we obtain $M_{500} = 3.7 (\pm 0.2) \times 10^{14}$\Msun, in agreement with the SZ-derived $M_{500}$ of $3.6 \times 10^{14}$\Msun, and LLS = 863~kpc, somewhat larger than our measurement of $\sim$660~kpc but within the uncertainties of the correlation.

\subsubsection{Southern Relic}

The southern relic is very long and curved with a convex outer edge roughly following the 16\arcmin\ diameter ring shown in Fig.~\ref{fig:fig2}. Its largest linear size (LLS) is at least 10\arcmin\ or 1.64~Mpc at the adopted distance, and it is located $\sim$1.3~Mpc from the cluster centre. The eastern end of the relic is less than 30\arcsec\ (82~kpc wide). This is consistent with a relic at a shock front, considering current formation theories / theoretical models \citep{Kang2017,Jones2023}. The large southern relic has a much smaller width than the less extended, northern relic. Within the wider central part of the southern relic, however, there is a prominent ridge and a wedge-like feature in the west, at an angle of $\sim$45\degr\ to the western end of the relic. The shape of the latter feature might be the signature of an inward moving shock \citep[see, e.g.,][]{Boess2023}. We measure a primary-beam corrected flux density of $\sim$10.5~mJy in the MeerKAT 1.3~GHz wide-band image, corresponding to a radio power of $P_{1300} \sim  7.1 \times 10^{23}$~W\,Hz$^{-1}$ at the adopted cluster distance. Apart from a small area within the wedge, the southern relic is not detected in RACS-low, neither is it detected in GLEAM or TGSS.

The average surface brightness of the southern relic is $\sim$16~$\mu$Jy\,beam$^{-1}$. Hence, we have found an example of a low surface brightness relic (about a factor of 10 below the mean surface brightness of the relics listed in \citet{Jones2023}) of which there are perhaps many more that have remained undetected by other surveys. In contrast, the northern relic has an average surface brightness of $\sim$55~$\mu$Jy\,beam$^{-1}$. The low surface brightness of the southern relic and its distance from the MeerKAT pointing centre ($38 \pm 3$ arcmin) does not allow for a reliable in-band spectral index measurement. Scaling the radio power to 150~MHz assuming a spectral index of $\alpha = -1$ \citep[see][their Fig.~6]{Jones2023} gives $P_{150} \approx 6.2 \times 10^{24}$~W/Hz. We obtain $M_{500} \approx 4.3 \times 10^{14}$\Msun, somewhat higher than the SZ-derived value of $3.6 \times 10^{14}$\Msun\ and LLS $\approx$ 950~kpc, much smaller than our measurement and about 2$\sigma$ from the correlation.

\subsubsection{Cluster centre}

\begin{figure} % Figure 8
\centering
  \includegraphics[width=8cm]{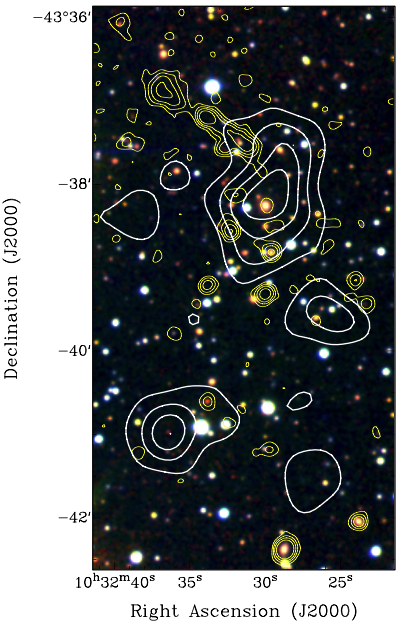}
\caption{MeerKAT 1.3 GHz radio continuum contours (yellow; 0.015, 0.03, 0.06, 0.12, 0.25 and 0.5 mJy\,beam$^{-1}$; resolution 7.7\arcsec) and 0.3--2 keV eRASS:3 X-ray contours (white; 0.002, 0.003, 0.004, and 0.005 cts\,s$^{-1}$; smoothed with a 30\arcsec\ Gaussian) of the PSZ2~G277.93+12.34 cluster centre overlaid onto an RGB colour image consisting of VHS $K_{\rm s}$-band (red), VHS $J$-band (green) and DSS2 $R$-band (blue). The radio-detected galaxy WISEA J103230.00--433815.4 \citep[\zph\ $\approx$ 0.158,][]{Bilicki2014}, which is located at the centre of the northern X-ray peak, is the likely BCG.}
\label{fig:psz2-center}
\end{figure}

The galaxy WISEA J103230.00--433815.4 \citep[\zph\ $\approx$ 0.158,][]{Bilicki2014} is located at the peak of the brightest eRASS:3 X-ray emission patch near the centre of the PSZ2 cluster (see Fig.~\ref{fig:psz2-center}). This is likely the brightest cluster galaxy (BCG); its redshift is similar to that of the host galaxy of PMN J1033--4335 (\zph\ $\approx$ 0.153) in the northern structure. Using the MeerKAT data we measure a radio flux density of 16~$\mu$Jy at 1.3~GHz, corresponding to a radio power of $\sim$1.1 $\times 10^{21}$ W\,Hz$^{-1}$. Around the BCG we note several other radio-detected likely cluster galaxies, incl. 2MASS J10322960--4338494 (WISEA J103229.72--433848.9). The southern X-ray peak in Fig.~\ref{fig:psz2-center} coincides with WISEA J103236.43--434058.4. \\

No radio halo is detected between the two relics in our MeerKAT 1.3~GHz images. Assuming a halo diameter of 3\arcmin\ (see Fig.~\ref{fig:relic-RGB}), we put an upper limit of 6~mJy on its diffuse flux density based on the rms in the area and the average surface brightness of the faint southern relic ($\sim$16~$\mu$Jy\,beam$^{-1}$), corresponding to a radio power of $P_{1300} \sim 4 \times 10^{23}$~W\,Hz$^{-1}$. Based on the scaling relation between the X-ray luminosity and the radio halo emission \citep[][and references therein]{Zhang2020}, we expect a radio halo power well below 10$^{23}$~W\,Hz$^{-1}$, in agreement with our upper limit.

\subsubsection{Other galaxies}

A prominent double-lobe radio (DRG) galaxy is included in Figs.~\ref{fig:fig1}--\ref{fig:relic-RGB}, located $\sim$15\arcmin\ east of the cluster centre. Its host galaxy is 2MASS J10334526--4335420 \citep[$z \approx 0.158$,][]{Gaia2022}, likely a cluster member. The DRG's radio lobes extend $\sim$4\arcmin\ North--South ($\sim$660~kpc), and its total flux density is $\sim$43~mJy in the RACS-low 888~MHz data and $40.4 \pm 4.2$ mJy in SUMSS 843~MHz data.

\subsection{eROSITA X-ray emission}

The PIB-subtracted and exposure-corrected eRASS:3 image reveals a disturbed patch of extended X-ray emission between the two radio relics, associated with the approximate PSZ2 G277.93+12.34 cluster center 
% (10:32:16.2, --43:39:58 $\pm$5.3\arcmin), 
and another patch around the NAT radio galaxy PMN~J1033--4335 (see Figs.~\ref{fig:fig2} \& \ref{fig:relic-RGB}). The radial extent of the hot cluster gas is $\sim$11.4\arcmin\ ($\sim$1.9~Mpc at $z \approx 0.158$) for the main structure, centred at 10:32:28.9, --43:38:26, and $\sim$9.2\arcmin\ ($\sim$1.5~Mpc at $z \approx 0.153$) for the northern structure, centred at 10:33:04.3, --43:35:23; see Table~\ref{tab:tab2}. The redshifts adopted above are for the respective central galaxies, WISEA J103230.00--433815.4 (\zph $\approx$ 0.158) and WISEA J103302.46--433508.5 (\zph $\approx$ 0.153). The available published redshifts for galaxies likely associated with the main and northern cluster structures are listed in Table~\ref{tab:tab3}. Since the PSZ2 G277.93+12.34 cluster lies just outside the eRASS legacy survey footprint, the two structures will not be included in the first eROSITA All-Sky Survey (eRASS:1) cluster catalog \citep{Bulbul-clusters2024}. More galaxy redshifts will become available from the second data release of the all-sky NOIRLab Source Catalog \citep{NSCDR2}. 

Through X-ray spectral analysis in the $0.3(0.8)-9.0$ keV band we obtained the following ICM properties: temperatures ($k_\mathrm{B}T$), normalizations ($norm$), and the X-ray luminosity ($L_{X,0.5-2.0~\mathrm{keV}}$). The reported luminosities (5.1 and 2.8 $\times\ 10^{43}$ erg\,s$^{-1}$ for the main and northern structures, respectively) are given at the source redshifts in the source frame energy band of 0.5--2.0 keV. The results are listed in Table~\ref{tab:tab4}, somewhat limited by our low counts. In the energy band used for the spectral fitting, we only have 680 and 420 counts in the main (m) and northern (n) structures, respectively, from all TMs combined. A first fit resulted in upper limits for the metallicities of $1.7~Z_\odot$ (m) and $1.5~Z_\odot$ (n). We proceeded with a fixed metallicity fit ($Z = 0.3~Z_\odot$), resulting in good agreement with the free metallicity fit but no improvement in the temperature constraints: $6.54_{-3.37}^{+4.02}$ keV (m) and $6.28_{-3.11}^{+20.93}$ keV (n). Deeper data, for instance, from a follow-up observation by \textit{XMM-Newton} or \textit{Chandra}, would help improve the ICM estimates of the temperatures and metallicities. We infer the $M_{500}$ mass by using the luminosity-mass ($L_X-M$) scaling relation from \cite{Lovisari_2015}
\begin{equation}
  \log(M_{500}/C_1) = a \cdot \log(L_X/C_2) + b,
\label{eq:scaling}
\end{equation}
with $a = 1.08 \pm 0.21$, $b = 0.18 \pm 0.18$, $C_1 = 10^{43}~h_{70}^{-2}$~erg~s$^{-1}$, and $C_2 = 2.0$ keV. The $R_{500}$ radii are calculated by assuming spherical symmetry and taking 500 times the critical density of the Universe at the assumed redshift. Using the SZ derived $M_{500}$ cluster mass ($3.6 \pm 0.6 \times10^{14}$\Msun, see Section~1) and $L_{X}$ from eROSITA (see Table~\ref{tab:tab4}), we find that PSZ2 G277.93+12.34 lies below the expected correlation by \citet{Pratt2009} and \citet[][their Fig.~22]{Planck2016}, but just within the 2$\sigma$ scatter. 
% Given the marginal SZ detection with Planck, its size and centre, it is likely that only the main structure is detected. 

\begin{table*} % Table 4
\centering
\begin{tabular}{cccccc}
\hline
 structure & $norm$ & $L_{X,0.5-2.0~\mathrm{keV}}$ & $M_{500}$ & $R_{500}$ & C-stat / d.o.f. \\ [5pt]
 & [$10^{-5}~\mathrm{cm^{-5}~arcmin^{-2}}$] & [$10^{43}$ erg\,s$^{-1}$] & [$10^{14}M_{\odot}$] & [arcmin] & \\ 
\hline
\hline
main (m) & $2.10_{-0.16}^{+0.26}$ & $5.08_{-0.67}^{+0.39}$  & $1.48_{-0.67}^{+0.65}$ & $4.61_{-0.69}^{+0.68}$ & 1828.61/2261 \\ [5pt]
\hline
north (n) & $1.79_{-0.21}^{+0.56}$ & $2.78_{-0.62}^{+0.30}$  & $0.85_{-0.38}^{+0.35}$ & $3.98_{-0.59}^{+0.54}$ & 1604.75/2047 \\ [5pt]
\hline
\end{tabular}
\caption{Derived eRASS:3 X-ray properties for the main and northern structures of the PSZ2~G277.93+12.34 galaxy cluster for the adopted redshifts of $z \approx 0.158$ (m) and $z \approx 0.153$ (n). The C-statistics per degree of freedom (d.o.f.) are given in the last column.}
\label{tab:tab4}
\end{table*}

% -- ORC J1027-4422 --
\begin{figure*} % Figure 9
\centering
 \includegraphics[width=8.2cm]{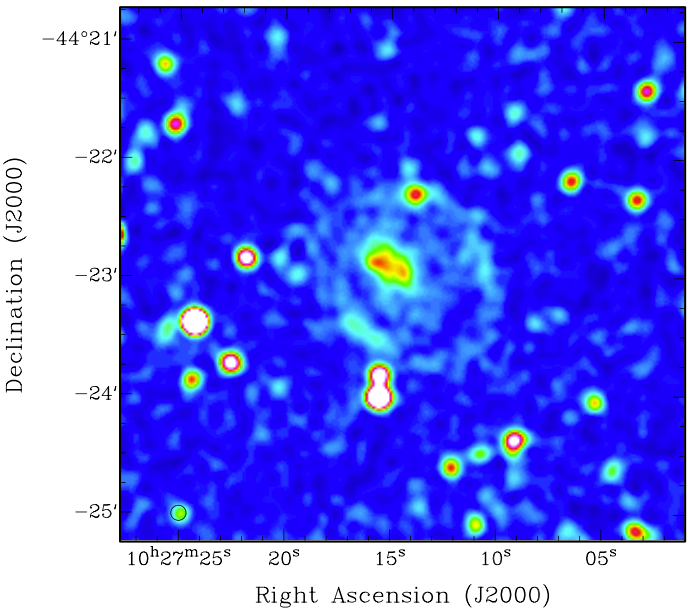}
 \includegraphics[width=8.1cm]{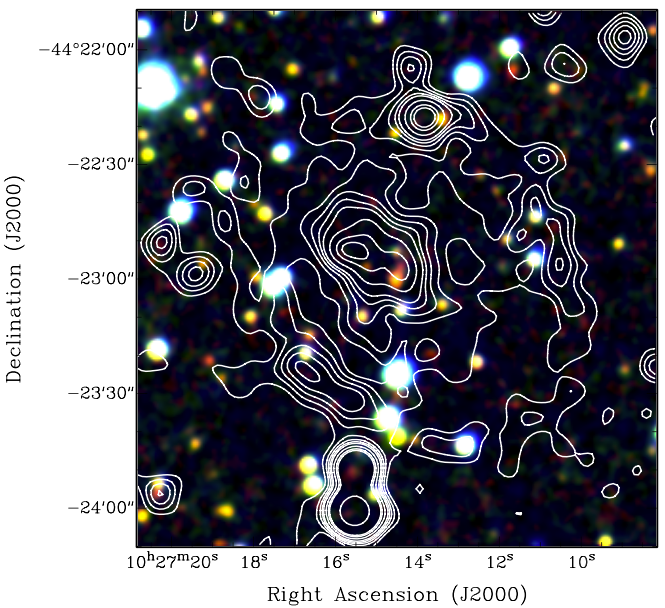}
\caption{ORC J1027--4422. {\bf --- Left:} MeerKAT 1.3 GHz radio continuum image. {\bf --- Right:} Zoomed-in RGB colour image consisting of VHS $K_{\rm s}$-band (red), VHS $J$-band (green) and DSS2 $R$-band (blue), all smoothed to 2\arcsec\ resolution, overlaid with MeerKAT radio contours at 3, 8, 13, 18, 23, 35, 50, 60 and 200 $\mu$Jy\,beam$^{-1}$. The displayed radio images are primary beam corrected and have an angular resolution of 7.7\arcsec.}
\label{fig:orcj1027-4422}
\end{figure*}
% Kvis RGB settings: 
% RED - Ks band: 362300 to 362900
% GREEN - J band: 83600 to 84200
% BLUE - DSS2 R band: 240000 to 300000

\section{ORC~J1027--4422}
% ORC J1027--4422 found on 8 Jun 2021

The discovery of Odd Radio Circles (ORCs) with ASKAP by \cite{Norris2021} and \cite{Koribalski2021}, prompted us to search for similar radio sources in the observed MeerKAT field. The three known single ORCs --- ORC J2103--6200, ORC J1555+2726 and ORC J0102--2450 --- have diameters of 70\arcsec\ -- 80\arcsec\ (300 -- 500~kpc) and are each centered on a prominent elliptical galaxy ($z \approx 0.2 - 0.6$). Their properties are summarised in Table~5. Our search resulted in one peculiar radio ring, named ORC~J1027--4422 (see Fig.~\ref{fig:orcj1027-4422}), which we analyse here. It consists of a very faint, patchy near-circular structure and a much brighter, elongated central area with two distinct radio peaks. The whole system, including the diffuse radio emission within the ring, has a total, primary-beam corrected flux density of $1.0 \pm 0.1$~mJy in the MeerKAT 1.3~GHz wide-band image. The ring has a slight ellipticity (diameter $\sim$ 90\arcsec\ $\times$ 100\arcsec, $PA \sim 45$\degr) and  varies in brightness. The two brightest ring sections, approximately south-east and north-west of the centre, are similar in morphology to double relics in the outskirts of galaxy clusters (like the one discussed in Section~3). ORC~J1027--4422 is centred at approximately $\alpha,\delta$(J2000) = 10:27:14.38, --44:22:56.7 (Galactic $l,b = 277.5\degr, +11.3\degr$). The elongated radio source in its central area ($\sim$0.4~mJy) has two distinct radio peaks (at 58 and 67 $\mu$Jy\,beam$^{-1}$) which are separated by $\sim$15\arcsec.

The centre position given above corresponds to the western radio peak, which may be associated with one or more distant galaxies detected in VISTA $J$- and $K_{\rm s}$-band images (see Fig.~\ref{fig:orcj1027-4422}, right): e.g., VHS J102714.34--442255.9, VHS J102714.48--442259.6, and VHS J102714.50--442252.5 \citep{McMahon2013}. The corresponding mid-infrared source, centered on VHS J102714.48--442259.6, is WISEA J102714.45--442258.9 with magnitudes (profile fit) of W1 = $15.46 \pm 0.04$, W2 = $15.25 \pm 0.08$, i.e., W1 -- W2 = $0.2 \pm 0.1$, and W3 $>$ 12.2, also detected in CatWISE2020 \citep{Marocco2021} as CWISE J102714.47--442259.5 with W1 -- W2 = 15.53 -- 15.41 = 0.12, typical for elliptical galaxies. No galaxies appear to be associated with the eastern radio peak, which could be a radio jet, but see further discussion below. Since neither spectroscopic nor photometric redshifts are available in this area we infer a redshift of $z \approx 0.3$ based on the red colour of the galaxy near the centre, which has an extinction-corrected $R$-band magnitude of $\sim$19.1, and comparison with large redshift catalogs. Within $\pm$0.1 of this magnitude the median redshift of $4.53 \times 10^6$ galaxies in \citet{Bilicki2016} is $0.23^{+0.05}_{-0.06}$.
% VHS J102714.34-442255.9: Ksp, Ksap3, Ksapc3 = 17.59, 18.02, 18.47; 
% K -> log M_star = 10.65, 10.5, 10.3
% L_K = D^2 10^(10-0.4*(Kmag-3.27))
% M_K/L_K = 1 Msun/Lsun
For a redshift of $z \approx 0.3$, the diameter of ORC~J1027--4422 is $\sim$400~kpc, similar to that of the three single ORCs (see Table~5). We use the central galaxy's $K_{\rm s}$-band magnitude to derive a stellar mass of $\sim$3 $\times 10^{10}$\Msun. Follow-up optical observations are under way.

\subsection{X-ray flux limit}
Using the eRASS:3 data described in Section~3.2, we derive an upper limit to the X-ray flux associated with ORC~J1027--4422. We adapt the Bayesian approach used by \cite{Willis2018} to obtain the X-ray aperture photometry for a sample of galaxy clusters observed with XMM-Newton to eROSITA. First we calculate the net source counts using independent aperture areas for the source and the background. For the source we choose a circular aperture area of 50\arcsec\ radius, and the eRASS:1 source catalogue is used to mask sources falling in either the aperture or background areas. To convert the estimated X-ray count-rate posterior probability density (ppd) of ORC~J1027--4422 into a flux density ppd, we use an energy conversion factor of $8.9 \times 10^{-13}$. The latter is computed using {\texttt XSPEC} \citep{Arnaud1996} and an {\texttt APEC} \citep[Astrophysical Plasma Emission Code,][]{Smith2001} model with a metallicity of 0.3~$\mathrm{Z_\odot}$, a temperature of 0.15~keV, redshift 0.3, $N_{\rm HI} \sim 8 \times 10^{20}$ from the \citet{HI4PI2016}, and the eROSITA response matrices extracted using the eSASS task \texttt{srctool} \citep{Brunner2022}. By multiplying the upper limit to the count rate of 0.0097 cts\,s$^{-1}$ by the above energy conversion factor we obtain a flux density upper limit of 8.6 $\times$ 10$^{-15}$~erg\,s$^{-1}$\,cm$^{-2}$ in the energy range of 0.2--10~keV. For comparison, the slightly deeper eROSITA Final Equatorial-Depth Survey (eFEDS) includes 542 candidate clusters and groups of galaxies ($z_{\rm median} = 0.35$) down to a flux limit of $\sim$10$^{-14}$\,erg\,s$^{-1}$\,cm$^{-2}$ in the 0.5--2~keV energy band within 1\arcmin\ \citep[see][]{Liu2022a}. Deeper X-ray observations of the area may reveal the hot intra-group gas associated with the ORC formation via galaxy mergers. For ORC~J1027--442 we estimate a luminosity limit of $3 \times 10^{42}$~erg\,s$^{-1}$. For comparison, this is a factor 10 fainter than the northern X-ray structure associated with the NAT radio galaxy.

\subsection{Spectral index estimates}
ORC~J1027--4422 resides 26.4\arcmin\ south-west of the MeerKAT pointing centre. To determine an approximate in-band spectral index for the elongated central emission of the system we measure the flux densities at both ends of the MeerKAT band. We obtain $1.2 \pm 0.1$~mJy at $\sim$888~MHz (compared to $1.4 \pm 0.4$~mJy in RACS-low) and $\sim$0.35~mJy at $\sim$1.6~GHz, which suggests a spectral index of $\alpha = -2.1 \pm 0.5$. Primary beam correction factors of 1.2 and 2.0, respectively, were applied.
% alpha = log(S1/S2) / log(nu1/nu2)
% --- ORC ---
% alpha = log(0.35/1.3) / log(1600/900)  = -2.3
% alpha = log(0.35/1.2) / log(1600/900)  = -2.1
% alpha = log(0.45/1.1) / log(1600/900)  = -1.6

While neither the central region nor the ring of ORC~J1027--4422 is detected in TGSS 150~MHz images, \citet{Intema2017} caution that TGSS is not sensitive to faint extended structures \citep[see also,][]{HW2017}. Deep low-frequency images are needed to detect the ORC and determine reliable spectral indices. Based on the TGSS rms we derive a 5$\sigma$ upper limit of 17.5~mJy\,beam$^{-1}$ for the relatively compact central region of ORC~J1027--4422, suggesting a spectral index of $\alpha > -1.7$ when compared with the MeerKAT 1.3~GHz wideband flux of 0.4~mJy. 
% alpha = log(0.4/17.5) / log(1300/150)  = -1.75
% alpha = log(1.2/17.5) / log(888/150)  = -1.50
We estimate a Mach number of $M > 2.0$ using $\alpha = - (M^2 +1)/(M^2 -1)$ and assuming diffusive shock acceleration \citep{Dolag2023}. For the adopted ORC redshift of $z \approx 0.3$, we derive a 150~MHz radio power of $P_{150} < 5.1 \times 10^{24}$~W\,Hz$^{-1}$. 
% Radio power calculation:
% P_radio [W/Hz] = 4 pi D_L^2 x S_radio
%  without k-correction (1+z)^(alpha-1)
%
For comparison, the median in-band spectral index of ORC~1's outer ring is --1.6, based on MeerKAT 1.3~GHz wide-band data \citep{Norris2022}, but varies considerably around the ring. A steeper index of --1.9 is found in the fainter regions. Using multi-telescope data from 88~MHz to 1.3~GHz, \citet{Norris2022} find a spectral index of $-1.4 \pm 0.1$ suggesting possible spectral steepening at higher frequencies, i.e. an ageing electron population.

\subsection{Formation scenarios}

\subsubsection{Galaxy merger shocks\,?}
The radio morphology of ORC~J1027--4422, in particular the two ring segments and the extended central emission, resembles that of a double-relic cluster with a central radio halo. Could the ring segments be shocks formed through galaxy mergers as recently proposed by \cite{Dolag2023}\,?  ORC~J1027--4422 is at least twice as distant and much smaller than the double radio relic associated with the PSZ2 G277.93+12.34 cluster, found in the same MeerKAT field and discussed in Section~3. Based on its size estimate, the $\sim$400~kpc ORC structure is more likely associated with a galaxy merger than a cluster merger. Similar ring/shell structures can be seen in \citet[][their Figs.~4 \& 7]{Dolag2023}. For a discussion of this scenario see Section~5.3 and Fig.~\ref{fig:merger-shocks}.

\subsubsection{A high-latitude SNR\,?}
ORC~J1027--4422 is different from previous single ORCs discussed by \citet{Norris2021} and \citet{Koribalski2021} in that it does not have a dominant central elliptical galaxy. It is therefore important to consider other formation scenarios. -- For example, could ORC~J1027--4422 be a young high-latitude ($b = +11.3\deg$) supernova remnant (SNR)\,? In that case, the extended radio source near its centre may be a pulsar wind nebula (PWN), similar to that in the Galactic SNR G09+01 \citep[][their Fig.~1]{Heywood2022GC}, or perhaps an accreting X-ray binary system, similar to the W50 / SS433 complex \citep{Dubner1998} or the Circinus X-1 source and its associated SNR \citep{Heinz2013,Coriat2019}. Both of these scenarios would not necessarily have optical counterparts. 
% No \Ha\ emission is visible in SuperCosmos images \citep{Parker2005}.
Future deep X-ray observations may allow detecting such PWN. ORC~J1027--4422 is a factor $\sim$2.4 smaller than the radio remnant around Circinus X-1 (distance $\sim$10 kpc) and much fainter in both radio and X-ray emission (undetected), suggesting that in this scenario it would be at least twice as distant. Known Galactic SNRs without optical or infrared counterparts are typically much larger than a few arcminutes. For example, DA~530 (G93.3+6.9) has a diameter of $\sim$27 arcmin, at least 44~pc for a minimum distance of 4.4~kpc, is highly polarised, and undetected in  optical/infrared emission \citep{landecker1999,booth2022}. Another example is the intergalactic SNR recently discovered near the Large Magellanic Cloud by \citet{Filipovic2022} which has a diameter of $\sim$200\arcsec\ ($\sim$48~pc assuming a distance of 50~kpc); no central radio emission is detected.

\section{Discussion}

\subsection{Radio relics and the disturbed ICM}

The galaxy cluster PSZ2 G277.93+12.34 is a disturbed system with an extended X-ray morphology and two prominent radio relics. The relics are separated by 16\arcmin\ (2.6~Mpc), located well beyond the cluster's detected X-ray emission (diameter = 11.4\arcmin\ or 1.9~Mpc). The X-ray emission of the ICM is elongated in the NE-SW direction and appears to contain two well-separated peaks (see Fig.~\ref{fig:psz2-center}). The direction of the presumed merger axis aligns with the direction of the radio relics likely resulting from the merger shock waves. The unusually large separation of the radio relics suggests that we see the merger after the first pericentre passage \citep{Zhang20}.

The measured sizes of the two radio relics (660 and 1640 kpc) differ by more than a factor two, suggesting that they result from an unequal mass merger. The lengths of the two relics depend on the mass ratio of the sub-clusters as well as the mass concentration of the sub-clusters prior to merger, as shown in hydrodynamical simulations by \citet{vanWeeren11_sim}. In these idealised simulations, the longer relic is located behind the more massive sub-cluster \citep[see also,][]{hoang2017}. The LLS of the southern relic is peculiar as it is about twice as large as the scaling relation between LLS and cluster mass would suggest \citep{Jones2023}. Strikingly, the widths of the two relics also differ significantly, with the shorter relic being much wider. The relics in the Sausage cluster, A3667, and A3376 also show this combination of short-roundish and long slim relics. This is not easy to explain since projection effects ought to affect both relics in a similar fashion. Double relics with different LLS and widths are also seen in the cluster A1240 and are discussed in \cite{Hoang2018}. We propose two reasons for the different appearance of the relics: --- (1) In unequal mass mergers the trajectory of the smaller cluster could be significantly bent by the gravity of the larger cluster such that the relics could be viewed under different angles. We suggest that the northern relic has been produced after the smaller subcluster has been deflected towards the line of sight. The thinner and longer southern relic, likely seen edge-on, has been produced inside the ICM of the smaller cluster when the latter first hit the ICM of the larger subcluster \citep[see also,][]{Lee2022}. --- (2) The northern relic is located in a different environment, perhaps related to the patch of X-ray emission in its vicinity. This could lead to a different Mach number or different turbulent velocities from the shock producing the southern relic. Deeper X-ray observations would reveal the detailed ICM distribution and help us better understand the merging geometry/configuration of the cluster \citep[see, e.g.,][]{Zhang2021}. 

The NAT radio galaxy's morphology is most likely caused by its fast outbound movement away from the cluster centre, possibly ejected by the merger shock that also caused the double radio relic. Bent tail galaxies is are often observed in clusters, but their geometries vary a lot due to the complexity of cluster merger evolution.

\subsection{Extended X-ray IC\,emission\,?}

The most likely explanation of the extended X-ray emission in the northern structure, around the NAT radio galaxy (PMN~1033--4335), is {\em thermal} emission from hot intracluster or intragroup gas. The X-ray center position of the northern structure (see Table~\ref{tab:tab2} and Fig.~\ref{fig:nat}) is $\sim$20\arcsec\ NE of the NAT's host galaxy. As the short exposure time of the eRASS:3 data do not allow us to unambiguously determine the origin of the X-ray emission, it is worth considering possibilities other than thermal emission. One may speculate that the northern X-ray detection is related to the NAT since it extends in the same direction as the bent radio lobes, however, it also extends in the opposite direction. The X-ray emission is also wider than the bent radio lobes, but does not extend into the northern radio relic. 

The synchrotron emission of PMN~J1033--4335 clearly shows the presence of an extended cloud of relativistic electrons. Photons from the cosmic microwave background (CMB) will necessarily inverse Compton (IC) scatter on those electrons, possibly giving rise to detectable extended non-thermal X-ray emission. Since $P_{\rm Compton} \sim \frac{u_{\rm phot}}{u_B}P_{\rm synchrotron}$, the expected X-ray luminosity from IC emission can be estimated by assuming a magnetic field energy density, $u_B$, the known CMB photon energy density, $u_{\rm phot}$, at the NAT redshift and the measured synchrotron power, $P_{\rm synchroton}$ \citep[e.g.,][]{2008MNRAS.386.1774E}. We compute the synchrotron and X-ray luminosities for a power-law energy distribution of cosmic ray electrons (CRe). Assuming that the radio spectral index reflects the slope of the CRe energy distribution for a homogeneous magnetic field distribution, we estimate the expected X-ray power of the NAT radio galaxy and compare it to our X-ray measurements. For a radio spectral index of $\alpha = -0.7$ , a flux density of 300~mJy at 888~MHz (ie, $P_{888}$ = $1.86 \times 10^{32}$~erg\,s$^{-1}$\,Hz$^{-1}$) and an assumed magnetic field of 3~$\mu$Gauss, we estimate the X-ray power in the 0.5 -- 2 keV range from IC~emission as $L_{\rm 0.5-2keV} = 1.1 \times 10^{41}$~erg\,s$^{-1}$. This is much less than the estimated X-ray power of the northern structure around the PMN~J1033--4335 radio galaxy from our eROSITA measurements, $L_{X,0.5-2.0~\mathrm{keV}} = 2.8 \times 10^{43}$~erg\,s$^{-1}$, given in Table~\ref{tab:tab4}. We note that approximately a third of the northern X-ray emission is possibly associated with the radio tails. 

To increase the expected X-ray power from IC emission, a steeper spectral index of the radio lobes and/or a smaller magnetic field strength would be required. Conceivably, aged plasma in the radio galaxy may reside in a weaker magnetic field due to the expansion of the lobes. This may cause the actual CRe energy distribution to be steeper than inferred from the radio spectrum assuming a homogeneous magnetic field. Assuming a CRe energy spectrum slope of --3.42, we estimate the X-ray luminosity of the PMN~J1033--4335 radio lobes to be $L_{\rm 0.5-2keV} = 1.43 \times 10^{42}$ erg\,s$^{-1}$. In addition, decreasing the assumed magnetic field strength to 1~$\mu$Gauss, which requires more CRe to be present to meet the radio luminosity, results in $1.62 \times 10^{43}$ erg\,s$^{-1}$ which is close to the X-ray luminosity of the northern structure. However, since only a fraction of the northern structure coincides with the lobes of PMN~J1033--4335, the IC emission could only explain part of the northern structure. It is evident that the X-ray emission does not trace well the structure of the lobes, although this might be an effect of the low sensitivity in the X-ray observations. Deeper observations are needed to determine the morphology of the X-ray emission in the northern structure. Overall, the IC emission may possibly explain part of the northern structure, however, with the existing data it is not possible to unambiguously decide on the origin of this X-ray emission. 

% P_150 [W/Hz] = 4 pi D^2_L S_150
% = 1.2x10^20 x D^2_L,Mpc x S_150,Jy
\begin{table*}
\centering
\begin{tabular}{cccccccccc}
\hline
 name & & $z$ & size & $S_{150}$ & log $P_{150}$ & log $M_{\star}$ & log $M_{500}$ & References \\
  & & & [kpc] & [mJy] & [W\,Hz$^{-1}$] & [\MMsun] & [\MMsun] \\
\hline
 ORC~J2103--6200 & ORC~1 & 0.55 & 520 & $38 \pm 6$ & 25.66 & 11.70 & 14.41 -- 14.58 & \citet{Norris2021,Norris2022} \\
 % D_L,Mpc = 3177.3
 ORC J1555+2726 & ORC~4 & 0.45 & 520 & $28 \pm 3$ & 25.45 & 11.26 & 13.20 -- 13.40 & \citet{Norris2021} \\
 % D_L,Mpc = 2497.7
 ORC J0102--2450 & ORC~5 & 0.27 & 300 & $17 \pm 1$ & 24.59 & 10.98 & 12.38 -- 12.53 & \citet{Koribalski2021} \\
 % D_L,Mpc = 1375.7
 ORC J1027--4422 & & $\sim$0.3 & 400 & $< 17.5$ & $< 24.70$ & 10.48 & 11.71 -- 11.72 & here \\
 % D_L,Mpc = 1552.7 
\hline
\end{tabular}
\caption{Properties of ORCs and their host galaxies. $S_{150}$ is the measured 150~MHz flux density and $P_{150}$ the derived 150~MHz radio power. Estimates of the host galaxy's stellar mass $M_{\star}$ are from \citet{Zou2019} for ORCs~1, 4, 5 and derived from the VISTA $K_{\rm s}$-band magnitude for ORC~J1027--4422. The $M_{500}$ estimates are based on the stellar-to-halo mass ($M_{\star} \propto M_{200c}$) relation from \citet{2020A&A...634A.135G}, where the range only reflects the uncertainty of the averaged value and does not include the scatter of individual systems. Conversion of $M_{200c}$ to $M_{500}$ is based on the fitting formulae given in \citet{2021MNRAS.500.5056R}. }
\label{tab:orc-properties}
\end{table*}

\subsection{Cluster and galaxy merger shocks}

\begin{figure} % Figure 10
\centering
    \includegraphics[width=8cm]{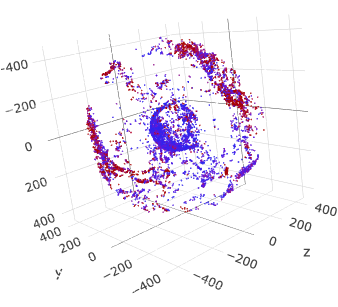}
    \includegraphics[width=8cm]{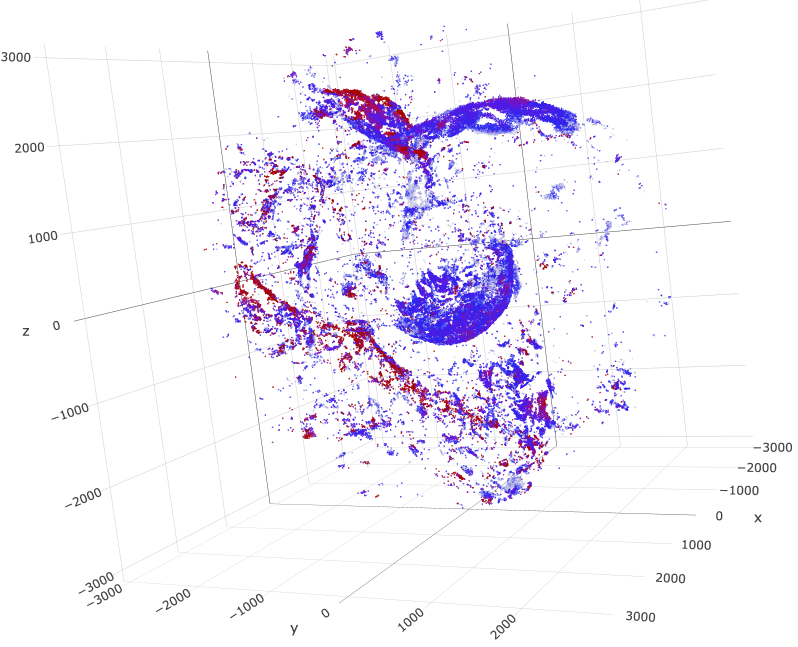}
\caption{Snapshots of galaxy and cluster merger simulations producing large-scale relics similar to those observed in ORC J1027--4422 (top panel) and in the PSZ2 G277.93+12.34 cluster (bottom panel). Interactive 3D visualisations of the galaxy merger shocks are available here \href{http://www.magneticum.org/complements.html\#Compass}{http://www.magneticum.org/complements.html\#Compass} presented in \citet{Dolag2023}. --- Inner and outer merger structures are visible at a range of Mach numbers. One prominent shock front shows a V-shape indicating that the center part is dragged into the cluster by infilling structures. The colours (blue to red) indicate Mach numbers from 2 to 5. }
\label{fig:merger-shocks}
\end{figure}

Cosmological, hydrodynamical simulations show that during the hierarchical formation of galaxies and galaxy clusters, large scale shocks are commonly present. As the merging of smaller entities to larger and larger structures is one of the main growth channels, such shocks are typically driven as part of the thermalization process of the hot atmospheres of these halos. To demonstrate the similarities of the morphology of these shock fronts with the observed radio emission, we analyzed two simulations from the COMPASS set, which are high resolution, zoom-in simulations of galaxies and galaxy clusters \citep{2011MNRAS.418.2234B}. These simulations where carried out with P-Gadget3, a modernized version of P-Gadget2 \citep{2005MNRAS.364.1105S}, that implements updated smoothed particle hydrodynamic (SPH) formulations regarding the treatment of viscosity and the use of kernels \citep{2005MNRAS.364..753D,2016MNRAS.455.2110B}, allowing a better treatment of turbulence within the circumgalactic medium and the intracluster medium. It also includes a formulation of isotropic, thermal conduction at 1/20th of the classical Spitzer value \citep{Spitzer1962} as well as an on-the fly shock finder \citep{2016MNRAS.458.2080B}. Fig.~\ref{fig:merger-shocks} shows a 3D visualisation of the detected shock geometry from the galaxy merger (top panel) and the cluster merger (bottom panel). Here the presented views are selected to closely represent ORC~1027--4422 as well as the a double relic similar to the one we found in the PSZ2~G277.93+12.34 cluster. Interactive plots which allow to explore the details in the 3D geometry are available at the link given in the figure caption. The simulated galaxy is the one presented in \citet{Dolag2023}, while the cluster is the {\it g6802296} halo, simulated at the same resolution than the one presented in \citet{2020MNRAS.494.4539Z}. The virial mass of the cluster at $z=0$ exceeds $10^{15}$\Msun, and the particle mass for dark matter (DM) and gas are $4.7\times10^6$\Msun\ and $8.9\times10^5$\Msun, respectively, and the softening for both, DM and gas particles is set to 0.69~kpc. Therefore, the cluster at redshift $z = 0$ is resolved with $\approx5\times10^8$ particles within the virial radius. The galaxy simulation uses 100 times smaller particle masses and a softening of 0.11~kpc.

While the morphologies and sizes of both ORCs and radio double relics suggest outward moving merger shocks on galaxy and cluster scales, respectively, the radio emission in the outer shells of the four known single ORCs is much brighter than expected from the cluster power-mass correlations \citep[e.g.,][]{deGasperin14,Jones2023}, diffuse shock acceleration \citep[DSA,][]{Hoeft2007} and in line with the simulations by \citet{Dolag2023}. The latter models currently do not reproduce the observed radio luminosities, due to the low, thermal energy content of the circumgalactic medium (CGM) at these distances. One possibility is that outwards-moving galaxy merger shocks are running into and compressing already existing (old) radio lobes or remnant lobes. This would make the radio shell emission much brighter (as needed) than the merger shocks just expanding into the CGM. The radio galaxy 3C\,442A possibly shows such a scenario as described by \citet{Worrall2007}.

\begin{figure} % Figure 11
\centering
    \includegraphics[width=8cm]{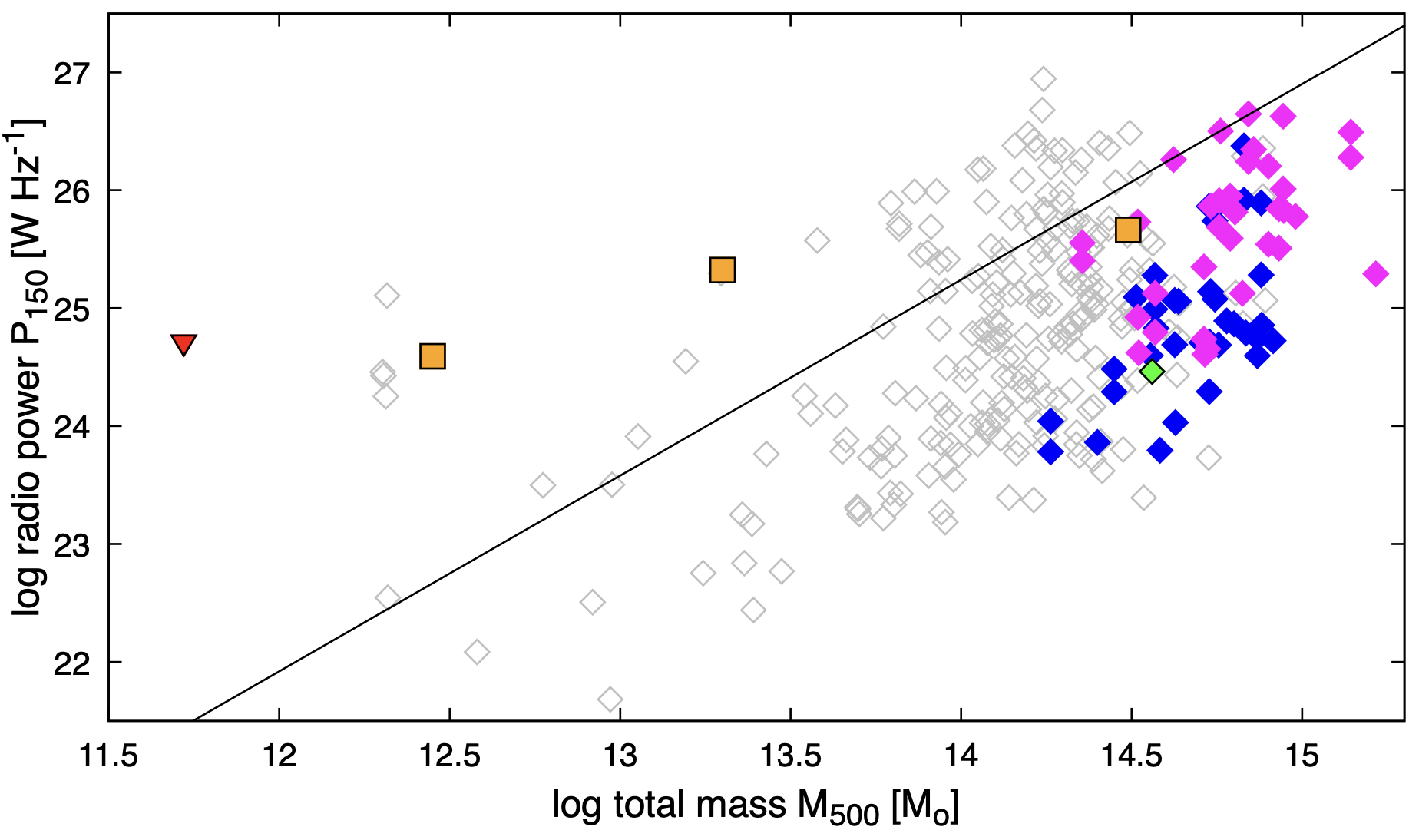}
\caption{Scaling relation of 150~MHz radio power ($P_{150}$) vs total mass ($M_{500}$) for cluster radio relics (incl. candidates) from \citet[][blue symbols]{Jones2023}, double relics from \citet[][pink symbols]{deGasperin14},  brightest cluster radio galaxies from \citet[][grey symbols]{Pasini2022} and ORCs (see Table~5). ORCs~1, 4, and 5 are indicated with orange squares and ORC~J1027--4422 with a red triangle. The northern relic of the PSZ2~G277.93+12.34 cluster (green symbol, see Section~3.1.2) follows the trend shown by \citet{Jones2023}. The black line indicates $P_{150} \propto M_{500}^{5/3}$ which corresponds to self-similar scaling assuming the same fraction of merging energy going into the radio power \citep[e.g.,][]{deGasperin14}. }
\label{fig:p150-m500}
\end{figure}

\section{Conclusions}

Radio relics are typically elongated, arc-shaped structures found in the outskirts of galaxy clusters ($>$10$^{14}$\Msun) with sizes up to $\sim$2~Mpc. Detected at $\sim$1~Mpc from the cluster centre, they trace merger-induced shock waves propagating through the ICM \citep[e.g.,][]{Ensslin1998}. For the special case of a merger in the plane of the sky, two relics are found on opposing sides of the merger axis (forming a partial circle), such that the shock fronts are seen approximately edge-on. Cosmological simulations of galaxy mergers reveal similar shock fronts of much smaller sizes \citep{Dolag2023}, providing a possible explanation for the formation of the recently discovered single odd radio circles \citep[ORCs,][]{Norris2021,Koribalski2021}. Here we report the radio and X-ray analysis of two new merger-related sources, a double radio relic surrounding hot cluster gas and a new odd radio circle, for which we present supporting simulations. Outward moving cluster and galaxy merger shocks are proposed as their respective formation processes (see Fig.~\ref{fig:merger-shocks}). These serendipitous discoveries were both made in the same deep, high-resolution MeerKAT 1.3~GHz radio continuum data, originally obtained to study the NGC~3263 galaxy group (Koribalski et al. 2024, in prep.). \\

The two relics, separated by 16\arcmin\ (2.6~Mpc at $z \approx 0.158$), are associated with the merging galaxy cluster PSZ2~G277.93+12.34 which has an SZ mass of $\sim$3.6 ($\pm$ 0.6) $\times$ 10$^{14}$\Msun. Complementary eRASS:3 data reveal two X-ray emission regions consisting of a near circular main structure in the cluster centre ($L_{X,0.5-2.0~\mathrm{keV}} \approx 5.1 \times 10^{43}~\mathrm{erg~s^{-1}}$) between the two radio relics and a northern structure surrounding the narrow-angle tail radio galaxy PMN~J1033--4335 ($L_{X,0.5-2.0~\mathrm{keV}} \approx 2.8 \times 10^{43}~\mathrm{erg~s^{-1}}$). Using the $L_X-M$ scaling relation we estimate $M_{500}$ masses of $\sim$1.5 and $\sim$0.9 $\times10^{14}M_\odot$ and $R_{500}$ radii of $\sim$4.6\arcmin\ and $\sim$4.0\arcmin\ for the main and northern cluster structures, respectively. PMN~J1033--4335 is located on the eastern side of the northern relic and, based on the location of its host galaxy and tail morphology, heading approximately north-east, i.e., away from the cluster centre. It was likely ejected outwards by cluster merger shocks that also resulted in the two radio relics. The northern relic has a linear extent of at least $\sim$0.66~Mpc and a surface brightness of $\sim$55~$\mu$Jy\,beam$^{-1}$. In contrast, the much thinner southern relic has a longer linear extent of $\sim$1.64~Mpc and a surface brightness of $\sim$16~$\mu$Jy\,beam$^{-1}$. Together the two relics form a partial circle, occupying at least $\sim$35\% of its circumference. The double relic morphology suggests a rare face-on orientation of the merger; only about two dozens such objects, including candidates, are currently known in the literature. Both the NAT and the northern relic are also detected in RACS-low at 888~MHz. Low-frequency imaging of the PSZ2 G277.93+12.34 cluster at both high resolution and high sensitivity would allow measuring the spectral indices of the relics, estimate their Mach numbers \citep{Hoeft2007} to learn more about the merger shocks, and possibly detect a radio halo. \\

The discovery of a peculiar radio ring, ORC~J1027--4422, of $\sim$90\arcsec\ diameter with an extended, double-peaked central emission area, makes it the 4th single ORC in the literature. The western peak, located at the ring centre, is likely associated with a galaxy or galaxy group. The eastern peak, possibly a radio jet, has no obvious optical counterpart. For the adopted redshift of $z \approx 0.3$ the ring diameter ($\sim$400~kpc) is similar to those of the previously discovered single ORCs. The morphology of ORC~J1027--4422 strongly resembles that of a double relic and radio halo, but a factor $\sim$5 smaller than typically found in galaxy clusters. The simulations by \citet{Dolag2023} suggest that the observed radio arcs / shells could be outwards moving merger shocks, occasionally forming during a massive galaxy merger. \\

To shed more light on the physical processes in action, we extend the radio power -- mass diagram (see Fig.~\ref{fig:p150-m500}) from relics and radio galaxies in clusters to ORCs surrounding massive early-type galaxies. Comparing to a simple, self-similarity model, assuming that the same fraction of merger energy is channeled into the radio emission \citep{deGasperin14}, we conclude that in ORCs the energy from the galaxy mergers is more efficiently channeled into the radio emission than in galaxy clusters. While for radio relics in the outskirts of galaxy clusters, less massive objects seems to be less efficient in channeling the merger energy into radio emission \citep[see discussion of possible origins for this in][]{deGasperin14}, galaxies hosting ORC systems are much more efficiently channeling the merger energy into radio emission. This indicates, that due to the more shallow potential well in the galaxy and small group like systems, AGN and star-formation activity can transport magnetic fields and cosmic rays much more efficiently to the outskirts and to larger distances, which then can be better lighted up (re-accelerated) and shine in radio when merger shocks travel through the IGM. The resulting central elliptical galaxy of ORC~J1027--4422 is estimated to have a virial mass of $\sim$10$^{12}$\Msun, making it by far the smallest system where potential merger shocks have been detected at radio frequencies. In addition, the extent of the radio emission exceeds $R_{500}$ by a factor of $\approx$1.7, which is the largest relative distance found so far and is significantly larger than the according ratio for the relics in PSZ2~G277.93+12.34. 

In a follow-up paper, we further explore the radio power -- mass diagram including recently discovered nearby radio shell systems that resemble ORCs: the Cloverleaf system \citep[][Koribalski et al. 2024b]{Dolag2023, Bulbul2024}, and the Physalis system \citep{Koribalski2024}. \\

Deep, wide-field radio continuum sky surveys with ASKAP at 0.7 to 1.8~GHz and $\sim$10\arcsec--30\arcsec\ resolution \citep[e.g.,][]{Koribalski2020,Norris2021}, which are well under way, will discover many clusters, either by detecting their extended radio halos and/or elongated relics, as well as more odd radio circles, essential for the study of their properties and formation mechanisms. ASKAP guest proposals will allow deeper follow-up studies of key target areas, very suitable for detailed galaxy cluster research as each ASKAP field spans 30 square degrees. 

% The last numbered section should briefly summarise what has been done, and describe the final conclusions which the authors draw from their work.

\section*{Acknowledgements}

We thank the referee for carefully reading the paper and suggesting several improvements. We also acknowledge valuable comments by Matthias Kluge, Natasha Hurley-Walker, Emil Lenc, Ned Taylor, Tadziu Hoffmann, Ildar Khabibullin and Tomoki Morokuma. \\

% MeerKAT acknowledgement
The MeerKAT telescope is operated by the South African Radio Astronomy Observatory, which is a facility of the National Research Foundation, an agency of the Department of Science and Innovation. 
% eROSITA
This work is based on data from eROSITA, the soft X-ray instrument aboard SRG, a joint Russian-German science mission supported by the Russian Space Agency (Roskosmos), in the interests of the Russian Academy of Sciences represented by its Space Research Institute (IKI), and the Deutsches Zentrum f\"ur Luft und Raumfahrt (DLR). The SRG spacecraft was built by Lavochkin Association (NPOL) and its subcontractors, and is operated by NPOL with support from the Max Planck Institute for Extraterrestrial Physics (MPE). The development and construction of the eROSITA X-ray instrument was led by MPE, with contributions from the Dr. Karl Remeis Observatory Bamberg \& ECAP (FAU Erlangen-N\"urnberg), the University of Hamburg Observatory, the Leibniz Institute for Astrophysics Potsdam (AIP), and the Institute for Astronomy and Astrophysics of the University of T\"ubingen, with the support of DLR and the Max-Planck Society. The Argelander Institute for Astronomy of the University of Bonn and the Ludwig Maximilians Universit\"at Munich also participated in the science preparation for eROSITA. The eROSITA data shown here were processed using the eSASS software system developed by the German eROSITA consortium. 
% VISTA
We also use infrared images obtained as part of the VISTA Hemisphere Survey, ESO Progam, 179.A--2010 (PI: McMahon). \\

BSK thanks Prof. Michael Kramer and the Max Planck Institut f\"ur Radioastronomie (MPIfR) in Bonn for their kind hospitality during many wonderful research visits where part of this paper was written. 
AV acknowledges funding by the Deutsche Forschungsgemeinschaft (DFG, German Research Foundation) -- 450861021. 
MB acknowledges support from the Deutsche Forschungsgemeinschaft under Germany’s Excellence Strategy -- EXC 2121 ”Quantum Universe” -- 390833306 and DFG Research Unit FOR 5195.
RJD is supported by BMBF grant 05A20PC4. 
KD acknowledges support by the COMPLEX project from the European Research Council (ERC) under the European Union’s Horizon 2020 research and innovation program grant agreement ERC-2019-AdG 882679 as well as by the Deutsche Forschungsgemeinschaft (DFG, German Research Foundation) under Germany’s Excellence Strategy -- EXC-2094 -- 390783311. 
XZ acknowledges financial support from the European Research Council (ERC) Consolidator Grant under the European Union’s Horizon 2020 research and innovation programme (grant agreement CoG DarkQuest No 101002585).

\section*{Data availability} 

The MeerKAT data used here are available through the SARAO Data Archive\footnote{https://www.sarao.ac.za/}, while the ASKAP data are available in the CSIRO ASKAP Science Data Archive (CASDA)\footnote{https://data.csiro.au/domain/casdaObservation}. The eRASS data from the first release are now also available online\footnote{https://erosita.mpe.mpg.de/dr1/}. All images used in this publication will be made available upon reasonable request to the lead author.

% The Acknowledgements section is not numbered. Here you can thank helpful colleagues, acknowledge funding agencies, telescopes and facilities used etc. Try to keep it short.

%%%%%%%%%%%%%%%%%%%%%%%%%%%%%%%%%%%%%%%%%%%%%%%%%%



\bibliographystyle{mnras}
\bibliography{master} 

%%%%%%%%%%%%%%%%%%%%%%%%%%%%%%%%%%%%%%%%%%%%%%%%%%

% Don't change these lines
\bsp	% typesetting comment
\label{lastpage}
\end{document}